\documentclass[twocolumn]{aastex701}
\usepackage{macros}
\usepackage{aas_macros}
\usepackage{amsmath}
\shortauthors{Barry et al.}

\begin{document}

\title{The origin of strong $\alpha$-element bimodalities in FIRE simulations of Milky Way-mass galaxies}

\author[orcid=0000-0002-1176-0078]{Megan Barry}
\affiliation{Department of Physics \& Astronomy, University of California, Davis, CA 95616, USA}
\email{mlbarry@ucdavis.edu}[show] 

\author[orcid=0000-0003-0603-8942]{Andrew Wetzel}
\affiliation{Department of Physics \& Astronomy, University of California, Davis, CA 95616, USA}
\email{awetzel@ucdavis.edu}

\author[orcid=0000-0003-3217-5967]{Sarah Loebman}
\affiliation{Department of Physics, University of California, Merced, CA 95343, USA}
\email{sloebman@ucmerced.edu}

\author[orcid=0000-0001-6380-010X]{Jeremy Bailin}
\affiliation{Department of Physics \& Astronomy, University of Alabama, Tuscaloosa, AL 35487-0324, USA}
\email{jbailin@ua.edu}

\author[orcid=0009-0007-3431-4269]{Hanna Parul}
\affiliation{LIRA, Observatoire de Paris, Universit\'e PSL, Sorbonne Universit\'e, Universit\'e Paris Cit\'e, CY Cergy Paris Universit\'e, CNRS, 92190 Meudon, France}
\email{hparul@crimson.ua.edu}

\begin{abstract}
One of the Milky Way's characteristic features is a strongly bimodal distribution of $\alpha$-process elements, such as Mg, at fixed [Fe/H] in stellar abundances.
We examine patterns in [Mg/Fe] versus [Fe/H] in FIRE-2 simulations of Milky Way-mass galaxies.
Out of 16 galaxies, 4 are capable of producing a strongly bimodal distribution.
In all four galaxies, the high-$\alpha$ population corresponds to an older, radially-compact, thick disk, and the low-$\alpha$ population corresponds to a younger, radially-extended, thin disk, similar to the MW.
The transition from high- to low-$\alpha$ took $0.3-1.2\Gyr$ and began $5.5-6.5\Gyr$ ago.
[Mg/Fe] decreased at relatively fixed [Fe/H], both in the galaxy overall and at fixed radii: Fe enrichment nearly balanced gas accretion (and therefore dilution), but Mg enrichment was weaker.
Importantly, this transition occurred during a period of relatively low gas fraction ($5-15\%$), immediately after a rapid decline in star formation (halving within a few hundred Myr), which caused an increase in Fe-producing white-dwarf supernovae relative to Mg-producing core-collapse supernovae.
Only one case coincided with a major merger coalescence.
We find similar trends in measuring stars by their current radius and by their birth radius, therefore, radial redistribution did not play a dominant role in the formation of a bimodality or its spatial dependence today.
Overall, in FIRE-2, strong $\alpha$-element bimodalities are relatively uncommon ($\sim25\%$), often not associated with a major merger, and arise primarily from a rapid decline in star formation during relatively low gas fraction.
\end{abstract}



\section{Introduction}
\label{sec:intro}

\begin{deluxetable*}{cccccccccccc}
\tabletypesize{\footnotesize}
\tablecaption{
Properties of each galaxy today.
}
\tablenum{1}
\tablehead{\colhead{Galaxy} & \colhead{$M_{\rm star}$} & \colhead{$R_{\rm star,50}$} & \colhead{$f_{\rm low-\alpha}$} & \colhead{$\Delta [\frac{\rm Mg}{\rm Fe}]$} & \colhead{$t_{\rm onset}$} & \colhead{$t_{\rm mid}$} & \colhead{$t_{\rm end}$} & \colhead{$\Delta t$} & \colhead{$t_{\rm disk \, onset}$}
\\ 
\colhead{} & \colhead{[$10^{10} \Msun$]} & \colhead{[kpc]} & \colhead{} & \colhead{[dex]} & \colhead{[Gyr ago]} & \colhead{[Gyr ago]}& \colhead{[Gyr ago]} & \colhead{[Gyr]} & \colhead{[Gyr ago]}
} 
\startdata
m12r & 3.9 & 1.37 & 0.39 & 0.06  & 5.5 & 4.6 & 3.9 & 1.6 & 7.9 \\
m12b & 8.5 & 1.56 & 0.46 & 0.05  & 5.8 & 5.2 & 4.6 & 1.2 & 7.4 \\
m12i & 5.7 & 1.38 & 0.46 & 0.03  & 6.5 & 6.25 & 6.2 & 0.3 & 6.1 (8.4) \\
m12q & 16.6 & 1.62 & 0.35 & 0.11  & 5.7 (6.7) & 5.4 & 5.1 & 0.6 (1.6) & 6.2 (8.6) \\
\enddata
\tablecomments{
$M_{\rm star,tot}$ is the total stellar mass of the galaxy.
$R_{\rm star,50}$ is the radius that contains 95\% of all stars.
$f_{\rm low-\alpha} = M_{\rm star,low-\alpha} / M_{\rm star,tot}$ is the mass fraction of low-$\alpha$ stars to total stars.
$\Delta \rm [Mg/Fe]$ is the ``gap width'' in [Mg/Fe] at fixed [Fe/H] between the peaks (modes) of the high- and low-$\alpha$ populations, averaged across the [Fe/H] with a well-defined bimodality.
$t_{\rm onset}$ is the age at which the low-$\alpha$ population began to rise significantly ($f_{\rm low-\alpha} \gtrsim 0.05$), $t_{\rm mid}$ is the age after which more than half of stars formed in the low-$\alpha$ population, and $t_{\rm end}$ is the age at which nearly all stars formed in the low-$\alpha$ population ($f_{\rm low-\alpha} \gtrsim 0.90$).
The values in parentheses for m12q indicate its first rise in $f_{\rm low-\alpha}$, which was followed by a prolonged period ($\approx 0.5$ Gyr) of $f_{\rm low-\alpha} \approx 0.2$, after which it began its second (primary) rise. We refer to these two $\alpha$ transition period onsets as $t_{\rm onset, 1}$ and $t_{\rm onset, 2}$.
$\Delta t$ is the duration of the $\alpha$ transition from $t_{\rm onset}$ to $t_{\rm end}$.
$t_{\rm disk \, onset}$ is the age at which a permanent disk formed, defined as $v_\phi / \sigma_{\rm tot} > 1$, as in \cite{McCluskey_2024}.
The value in parentheses for m12i indicate the first time the disk formation criteria was met; the disk was destroyed and reformed near the time of the $\alpha$ transition period.
For reference, some MW values are $M_{\rm star, tot} \approx 5 \times 10^{10} \Msun$, $R_{\rm half-light} = 5.75 \kpc$, $f_{\rm low-\alpha} \approx 0.8$, and $\Delta [\frac{\rm Mg}{\rm Fe}] = 0.11$ \citep{Bland_Hawthorn_2016,Vincenzo_2021,Lian_2024}. We note that the R50 values are smaller than the MW and probably reflect higher bulge fractions in FIRE galaxies.
}
\label{tab:galaxies}
\end{deluxetable*}

The Milky Way (MW) is our best-studied galaxy, with plentiful recent data \citep{Bland_Hawthorn_2016}.
Because stars (nearly) preserve the elemental abundances from their birth interstellar medium (ISM) in their atmospheres, many works use elemental abundance patterns in stars today to reconstruct the MW's history \citep[for example][]{Belokurov_2013, Helmi_2020, Deason_2024, Imig_2025}. Two abundances are particularly important: iron, the primary product of white-dwarf supernovae (WDSN, also known as Type Ia), and the $\alpha$-process elements (O, Ne, Mg, Si...), the primary products of core collapse supernovae (CCSN, also known as Type II).
In general, these abundances increase over time in a galaxy as stars form and enrich the ISM, but the different rates at which different stellar processes occur cause the relative abundances of different elements to change within a galaxy.
Therefore, by analyzing the abundances of different elements locked up in stellar atmospheres today across large populations within a galaxy like the MW, we can identify coherent populations and events in the MW's history.

One common way to probe a galaxy's history is via the ratio [$\alpha$/Fe] versus [Fe/H] of stars across the galaxy. [Fe/H] provides a (rough) sense of when a star formed, with stars generally expected to increase in Fe enrichment over time \citep{Tinsley_1979}.
[$\alpha$/Fe] provides a sense of the relative contribution of each supernova type throughout the galaxy's history up to the point that the star formed.
Higher [$\alpha$/Fe] indicates dominance of CCSN, and lower [$\alpha$/Fe] indicates dominance of WDSN  \citep[for example][]{Kobayashi_2020}.

CCSN occur when a massive star ($\gtrsim 8 \Msun$) exhausts its available fuel, leading to an implosion in the core. Prior to the SN, the core is surrounded by different layers, each undergoing different fusion processes that ultimately originated from helium fusion, hence, $\alpha$ element.
In practice, astronomers either choose one element to represent $\alpha$ (such as Mg), or average across many $\alpha$ elements in a stellar atmosphere.
These $\alpha$ elements comprise the majority of metals that CCSN enrich, although CCSN also produce some Fe \citep{Timmes_1995, Kobayashi_2006}. Because massive stars have short lifetimes ($\lesssim 40 \Myr$), all CCSN occur soon after star formation began, so the galaxy's oldest stars tend towards high [$\alpha$/Fe] at low [Fe/H].

Multiple processes can produce WDSN, including the thermonuclear detonation of a WD via accreted material from a companion star pushing it over the Chandrasekhar limit, and WD-WD mergers.
Fe constitutes the majority of metals that WDSNe release, although WDSNe produce a few $\alpha$ elements \citep{Kobayashi_2009, Kobayashi_2020}. Being the inert cores of longer-lived, lower-mass stars, WDs take time to form within a stellar population and continue to form (nearly) indefinitely. Therefore, WDSNe have a delayed onset ($\gtrsim 40 \Myr$ or much longer) and can continue to occur for many Gyr after a stellar population forms. Thus, a star higher in [Fe/H] and lower in [$\alpha$/Fe] indicates increasing dominance of WDSNe. Over time, at higher [Fe/H], stars tend towards lower [$\alpha$/Fe].

Therefore, we expect the oldest stars within a galaxy to form an initial plateau in [$\alpha$/Fe] at low [Fe/H], representing the [$\alpha$/Fe] ratio of CCSN products, and once a significant number of WDSNe start to occur, [$\alpha$/Fe] to decrease over time with increasing [Fe/H]. Indeed, almost all simulations and analytic models produce such behavior \citep{Brook_2012, Stinson_2013, Martig_2014a,Weinberg_2023}, and observations of lower-mass galaxies around the MW display this pattern \citep{Kirby_2011,Nidever_2020,Hasselquist_2021, Fernandes_2023}.

While abundance measurements of the MW follow the general trend of older, lower [Fe/H] stars having higher [$\alpha$/Fe] and younger, higher [Fe/H] stars having lower [$\alpha$/Fe], an unexpected feature in its abundance distribution is the presence of two distinct populations or tracks: a ``high-$\alpha$'' population that roughly follows the predicted pattern and corresponds to the thick disk, and a separate ``low-$\alpha$'' population that corresponds to the thin disk \citep[for example][]{Nidever_2014, Hayden_2015}. Each population covers a wide range of [Fe/H], with a ``gap'' at intermediate [Mg/Fe]. The origin of these separate populations remains debated. Since the first evidence in the late 1990s and early 2000s \citep{Fuhrmann_1998, Prochaska_2000,Reddy_2006}, this $\alpha$ bimodality of the MW has been confirmed across many surveys \citep[for example][]{Adibekyan_2011, Anders_2014, Nidever_2014, Hayden_2015, Vincenzo_2021}.

One outstanding question is whether such an $\alpha$ bimodality is common or rare in galaxies of a given mass.
Recent measurements of M31, the closest massive disk galaxy to the MW, do not show clear evidence for a distinct low-$\alpha$ population \citep{Nidever_2024}.
Astronomers have measured [$\alpha$/Fe] versus [Fe/H] patterns in stars for about a dozen low-mass satellite galaxies around the MW, including the moderately-disky LMC, but none show a clear $\alpha$ bimodality like the MW \citep[][]{Vargas_2013, Nidever_2020, Hasselquist_2021, Fernandes_2023}.
Thus, current observational evidence suggests strong $\alpha$ bimodalities are rare, but statistics are limited.
Cosmological simulations predict that a subset of such low-mass galaxies should show some multi-modality in [$\alpha$/Fe] versus [Fe/H] \citep{Patel_2022}, if one can measure stellar abundances to sufficient precision.

Observations show a strong link between the high- and low-$\alpha$ populations and the thick and thin disks of the MW \citep{Lee_2011, Bovy_2012c, SilvaAguirre_2018, Khoperskov_2024}.
The high-$\alpha$ thick disk is older and more radially compact, while the low-$\alpha$ thin disk is younger and more radially extended.
Many theoretical and observational works indicate that disk galaxies like the MW typically form radially ``inside-out'' and vertically ``upside-down'', because earlier stars formed radially closer to the galaxy center and vertically farther from the galactic plane \citep[for example][]{Bird_2013, Bird_2021, Yu_2023, Graf_2024}.
Thus, even if the MW did not have an $\alpha$ bimodality, we would still expect older stars to be in a thicker disk and higher in [$\alpha$/Fe], while younger stars are in a thinner disk and lower in [$\alpha$/Fe].

Many works have proposed different explanations for the origin of the $\alpha$ bimodality in the MW.
The earliest, called the ``two-infall model'' \citep{Chiappini_1997}, states that the high- and low-$\alpha$ populations originated from two separate star formation events, each triggered by a rapid deposition of Fe-poor gas that significantly decreases the galaxy's overall [Fe/H], which could be triggered by a gas-rich merger.
This model predates the discovery of the now-well-known Gaia-Sausage-Enceladus (GSE) merger \citep{Helmi_2018, Belokurov_2018}, and many works proposed that this event played a role in forming the MW's high- and low-$\alpha$ populations \citep{Vincenzo_2019, Palla_2020, Spitoni_2020}. 
In these scenarios, the merger either quenches or strongly perturbs the early high-$\alpha$ disk and then, after a phase of metal-poor gas accretion, reignites star formation along a second,  distinct low-$\alpha$ track.
In some simulations, clear $\alpha$ bimodality most often follows an early merger- or inflow-triggered high-SFR phase and a subsequent lull, with later, comparatively metal-poor accretion helping to build a younger low-$\alpha$ disk \citep{Mackereth_2018, Grand_2018,Beane_2025}.
These studies agree with the two-phase star formation history aspect of the two-infall model, but do not require a merger-driven, galaxy-wide [Fe/H] reset; rather, SFR-driven enrichment with modest, localized dilution is sufficient to produce a bimodality.
Other simulations suggest that more gradual accretion, not associated with a specific satellite, may be the dominant mechanism \citep{Khoperskov_2021, Agertz_2021}.
Other simulations suggest that extended, metal-poor CGM inflows not associated with a single satellite can generate a bimodality by forming low-$\alpha$ stars in the outer disk while the inner disk experiences a lull, via feedback-driven supply cycles \citep{Khoperskov_2021} or misaligned outer-disk accretion that builds a younger thin component alongside an older high-$\alpha$ component \citep{Agertz_2021}.

Some analytic models suggest dynamical explanations for the bimodality, or at least its spatial distribution in the MW today.
\cite{Schonrich_2017} ties the bimodality to inside-out growth driving a downward evolution in [$\alpha$/Fe] and radial abundance gradients, with radial redistribution apparently enhancing the effect.
Similarly, \cite{Sharma_2021} posits that a rapid, post-burst decline in [$\alpha$/Fe] with only modest change in [Fe/H] produces two tracks that become observationally distinct once radial redistribution mixes stars from different radii.

Some simulations attribute the bimodality to effects of clumpy, spatially-distinct star formation \citep{Clarke_2019, Amarante_2020, Garver_2023}.
Large-volume simulations have so far have shown strong bimodalities to be uncommon-to-rare among MW-mass disk galaxies: for example, in EAGLE, only $\sim5\%$ of MW-mass analogues show a pronounced bimodality \citep{Mackereth_2018}.
The Auriga simulations exhibits a wide diversity with many systems lacking a strong bimodality, although some degree of bimodality is not uncommon \citep{Orkney_2025}. \cite{Khoperskov_2021} also notes that incidence can depend sensitively on the strength of feedback cycles. The NIHAO simulations \citep{Buck_2020} found some degree of bimodality to be relatively common ($\sim 50\%$), emerging in systems with a pre-existing high-$\alpha$ disk that undergoes a gas-rich merger.
Overall, different cosmological galaxy simulations have shown that a strong bimodality can arise, though there is no single unifying explanation for producing an $\alpha$ bimodality.

We use the FIRE-2 simulations \citep{Hopkins2018,Wetzel_2023} to examine the origin of MW-mass galaxies with a ``strong'' bimodality in $\alpha$ elements, in which we can completely distinguish the high- and low-$\alpha$ populations with minimal overlap in [$\alpha$/Fe]-[Fe/H].
We build on recent work from \cite{Parul_2025}, who examined the fiducial set of MW-mass galaxies in FIRE-2, including both bimodal and non-bimodal galaxies. \cite{Parul_2025} classified a majority of the fiducial FIRE-2 galaxies as ``weakly bimodal'' or ``bimodal'', with one ``strongly bimodal'' galaxy, and linked the presence of this feature to late, Fe-poor inflows. Our work here includes both the strongly bimodal galaxy examined in \cite{Parul_2025}, m12b, as well as 3 more strongly bimodal galaxies from FIRE-2 simulations with variations on resolution and UV background model. We examine properties of the high- and low-$\alpha$ populations in each strongly bimodal galaxy, noting their common features and properties related to their star formation histories, such as star formation rate (SFR), supernova (SN) rates, and gas fraction, and we describe the galactic environment at the time of transition from high to low $\alpha$. We also characterize this ``$\alpha$ transition period'' in terms of its onset time and duration.

\section{Methods}
\label{sec:style}

\subsection{FIRE simulations}
\label{subsec:FIRE}

\begin{figure*}[t]
\centering
\includegraphics[width = \linewidth]{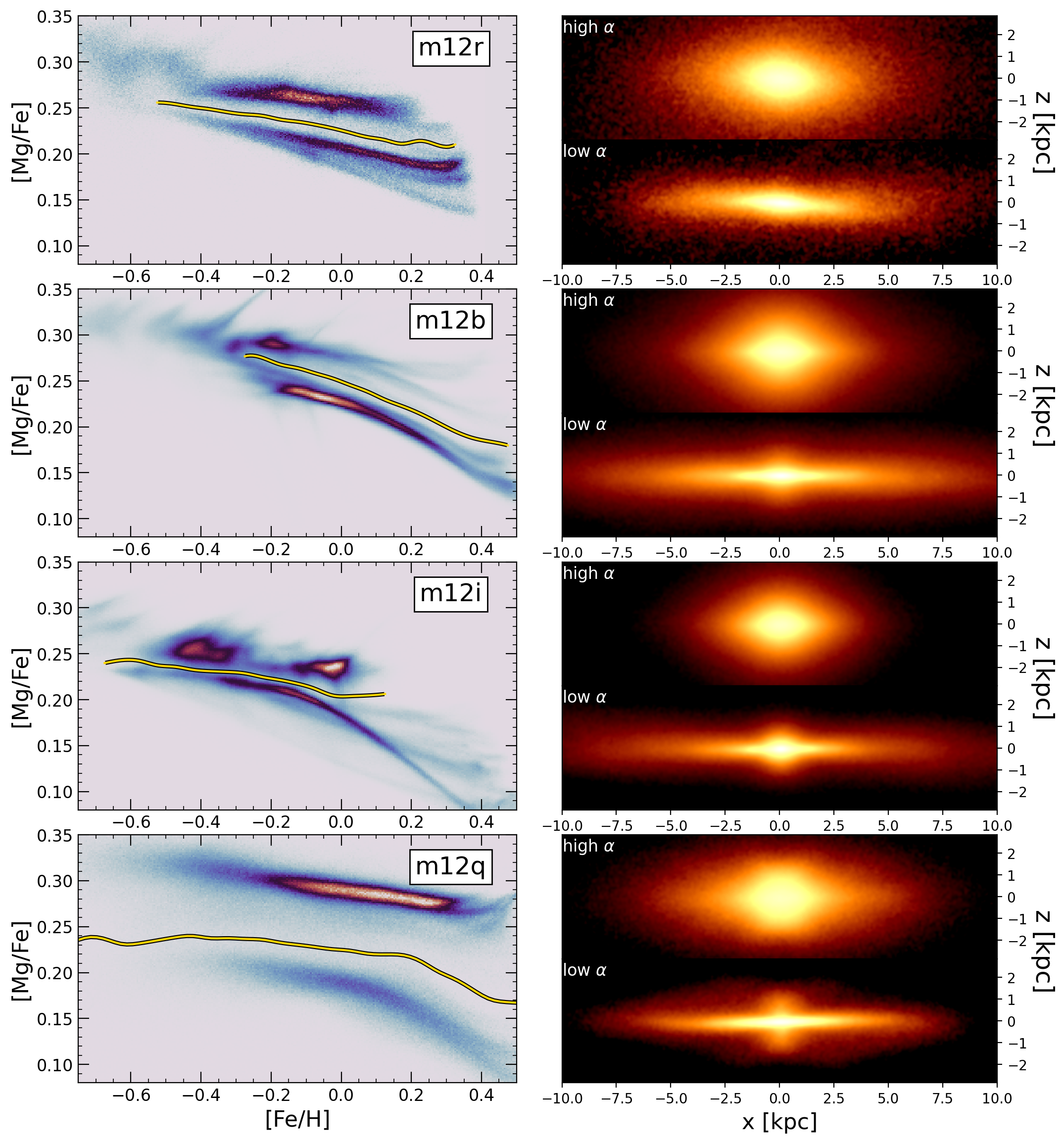}
\caption{
\textbf{Left}: 2-D histogram of [Mg/Fe] versus [Fe/H] today, color-coded linearly by the mass in stars, for the 4 MW-mass galaxies in the FIRE-2 simulations with a strong bimodality in $\alpha$-elements.
The oldest stars populate the upper-left region, while the youngest stars populate the lower region.
Each yellow curve delineates the boundary between well-defined high- and low-$\alpha$ populations (see Section~\ref{subsec:define}). We include stars to the left of this region as high-$\alpha$ and stars to the right of this region as low-$\alpha$.
\textbf{Right}: Images of stars today, color-coded by logarithmic mass, separately for the high- and low-$\alpha$ populations. All 4 galaxies are similar to the MW, in terms of the spatial correlations of high-$\alpha$ stars with the older, radially compact, thick disk (average age $t_{\rm birth} = 8.5 \Gyr$, average $R_{\rm star,95} = 7.5 \kpc$, and average $Z_{\rm star,95} = 1.6 \kpc$), and of low-$\alpha$ stars with the younger, radially extended, thin disk (average age $t_{\rm birth} = 3.1 \Gyr$, average $R_{\rm star,95} = 10.3 \kpc$, and average $Z_{\rm star,95} = 1.0 \kpc$).
}
\label{fig:disks}
\end{figure*}

We analyze MW-mass galaxies from the FIRE-2 cosmological zoom-in simulations \citep{Hopkins2018}. We generated each simulation using \textsc{Gizmo} \citep{Hopkins2015}, which models $N$-body gravitational dynamics with an updated version of the \textsc{GADGET-3} TreePM solver \citep{Springel2005}, and hydrodynamics via the meshless finite-mass method. FIRE-2 incorporates a variety of gas heating and cooling processes, including free-free radiation, photoionization and recombination, Compton, photo-electric and dust collisional, cosmic-ray, molecular, metal-line, and fine-structure processes, including 11 elements (H, He, C, N, O, Ne, Mg, Si, S, Ca, Fe).
FIRE-2 includes a spatially-uniform, time-dependent cosmic ultraviolet (UV) background from \cite{Faucher2009}, in which hydrogen reionization occurs at $z \approx 10$.
(One galaxy, m12i, uses the updated UV background from \cite{Faucher_2020}, as we describe below.)
Each simulation consists of dark matter, stars, and gas.
Star formation occurs in gas cells that are self-gravitating, Jeans-unstable, cold ($T < 10^{4} \K$), dense ($n > 1000 \cci$), and molecular as in \cite{krumholz}. Once formed, star particles undergo several feedback processes, including core-collapse and white-dwarf supernovae, continuous stellar mass loss, photoionization and photoelectric heating, and radiation pressure.

We generate cosmological initial conditions for each simulation at $z \approx 99$, within periodic boxes of length $70.4 - 172 \Mpc$ using the MUSIC code \citep{MUSIC}. We use a flat $\Lambda$CDM cosmology, with slightly different parameters across the simulations: $h = 0.68 - 0.71$, $\Omega_{\rm \Lambda} = 0.69 - 0.734$, $\Omega_{\rm m} = 0.266-0.31$, $\Omega_{\rm b} = 0.0455 - 0.048$, $\sigma_{8} = 0.801 - 0.82$, and $n_{\rm s} = 0.961 - 0.97$, broadly consistent with \cite{Planck}. We save 600 snapshots from $z = 99$ to $0$, with typical spacing $\lesssim 25 \Myr$.

\begin{figure*}[t]
\centering
\includegraphics[width = \linewidth]{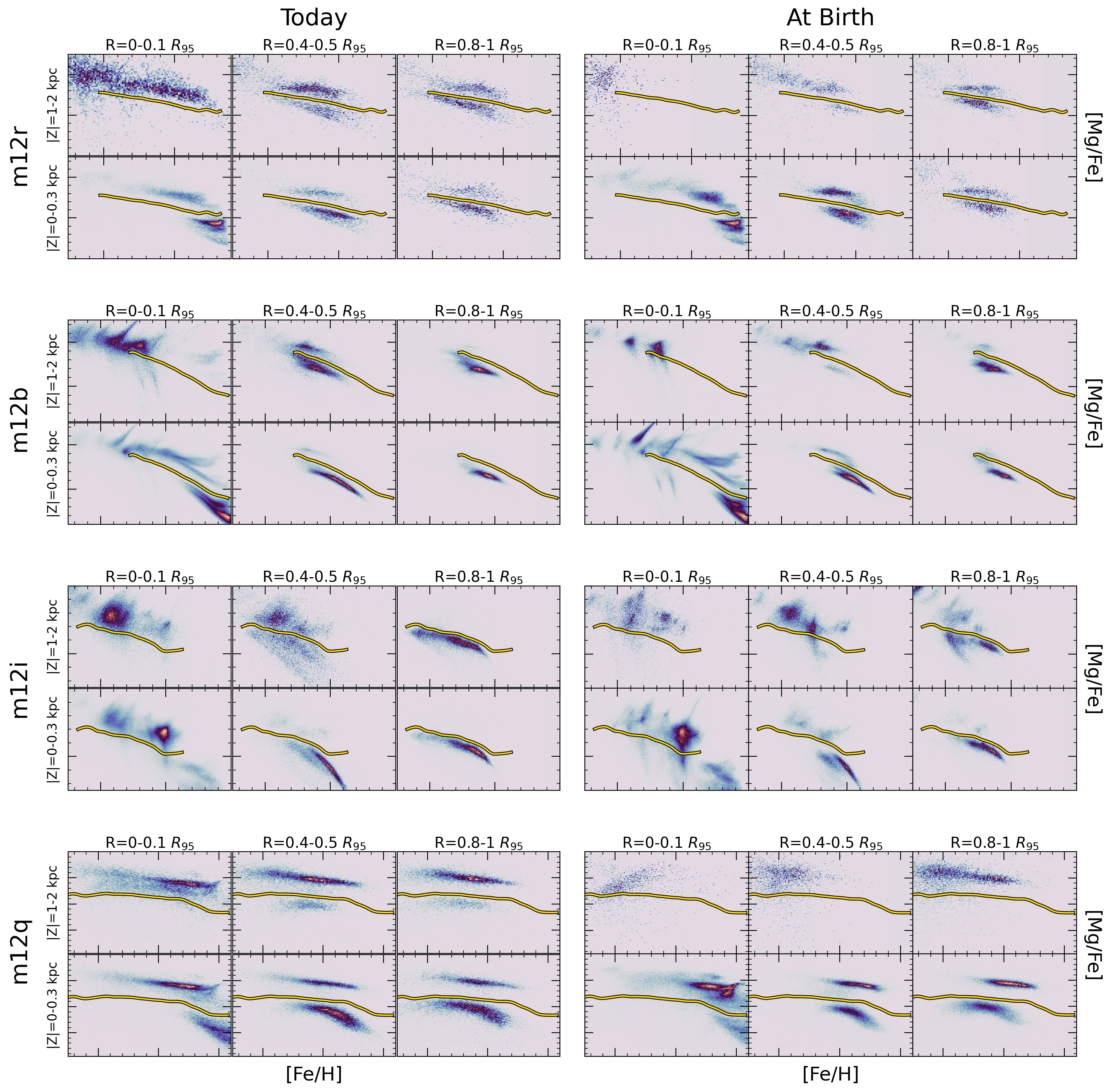}
\caption{
[Mg/Fe] versus [Fe/H] of stars at different spatial locations ($R$ and $Z$) in the 4 FIRE-2 galaxies with a strong bimodality in $\alpha$-elements, based on stellar positions today (left column) and at birth (right column). Both columns are measured in reference to $R_{\rm star,95}$ measured today.
The trends for stars today qualitatively agree with the MW \cite[for example][]{Anders_2014, Nidever_2014, Hayden_2015, Vincenzo_2021}, in terms of low-$\alpha$ stars dominating at small $Z$ and large $R$, and high-$\alpha$ stars dominating at large $Z$ and small $R$.
Similar to the MW, low-$\alpha$ stars have a negative radial gradient in [Fe/H].
The main exception is m12r, the lowest-mass galaxy, which shows significant high- and low-$\alpha$ populations across the entire disk.
\textit{As the right column shows, these trends for stars were largely present at birth, so they were not caused by radial redistribution of stars after birth.}
The main difference is that at birth there were fewer stars at high $Z$, which arises from dynamical heating of stars over time.
}
\label{fig:hayden}
\end{figure*}

Critical for this work, these FIRE-2 simulations explicitly model the subgrid diffusion/mixing of elemental abundances in gas that occurs via unresolved turbulence, assuming that the largest unresolved eddies dominate the mixing \citep{Su_2017, Escala_2017, Hopkins2018}.
Earlier FIRE-1 and some FIRE-2 simulations did not incorporate this metal mixing \citep[for example,][]{Ma_2017}, so they dramatically overpredicted the scatter in elemental abundances.
\cite{Escala_2017} showed that including this allows FIRE-2 to match observed distributions of stellar metallicities in low-mass galaxies.
\cite{Bellardini_2022} showed that this metal mixing is critical for matching the azimuthal variations in ISM abundances in MW-mass galaxies today, but also that it does not significantly affect radial or vertical gradients, because FIRE resolves the mixing relevant to those (larger) scales.

We examine MW-mass galaxies from two suites of simulations. The first is the \textit{Latte} suite of individual MW-mass halos \citep[introduced in][]{Wetzel2016}, which have dark-matter halo masses of $M_{\rm 200m} = 1 - 2 \times 10^{12} \Msun$ and no neighboring halos of similar or greater mass within at least $\approx 5 \Rthm$, where $\Rthm$ is the radius within which the density is 200 times the mean matter density of the universe. For the high-resolution simulations in this suite, gas cells and star particles have initial masses of $7070 \Msun$, while dark-matter particles have a mass of $3.5 \times 10^{4} \Msun$. The gravitational force softening lengths are 40 pc for dark matter and 4 pc for star particles (comoving at $z > 9$ and physical thereafter). In this work, we also examine some lower-resolution versions, in which all particles are $8 \times$ more massive and softening lengths are $2 \times$ larger.
Specifically, gas cells and star particles have initial masses of $57,000 \Msun$, while dark-matter particles have a mass of $2.8 \times 10^{5} \Msun$, with gravitational force softening lengths of 80 pc for dark matter and 8 pc for star particles.
All simulations use adaptive gravitational softening for gas, which matches the hydrodynamic smoothing, down to 1 pc.
We also examined galaxies from the ELVIS on FIRE suite of Local Group-like MW+M31 halo pairs \citep[introduced in][]{GK2019}. 
However, all 4 cases of a strong bimodality in $\alpha$-elements are from the \textit{Latte} suite, so our analysis focuses on those.

For white-dwarf supernovae (WDSN), we use the delay-time distribution (DTD) from \cite{Mannucci_2006}, with the specific event rate
\begin{equation}
\begin{split}
1.6\times10^{-5}\exp\left[-\frac{1}{2}\left(\frac{\tau-50}{10}\right)^2\right]
\\
\quad + 5.3\times10^{-8}\,\Msun^{-1}\,\mathrm{Myr}^{-1}
\end{split}
\end{equation}
where $\tau$ is the stellar age in Myr, and WDSN begin to occur at $\tau = 37.53 \Myr$. The first term represents a prompt component, and the second term represents a delayed, constant component that starts to dominate at $\tau \approx 100 \Myr$.
Recent work suggests that a power-law DTD, as in \cite{Maoz_2017}, may be more accurate \citep{Mohapatra_2025}.
\cite{Gandhi_2022} explored the use of the DTD from \cite{Maoz_2017} in FIRE-2 simulations: because it leads to $\approx 80\%$ more WDSN (depending on age), it lowers the [Mg/Fe] of stars and provides somewhat better agreement with the MW.
The newer FIRE-3 model switched to this DTD \citep{Hopkins_2023}, and in future work, we will examine the role of different assumed WDSN DTDs, but in this work, we use only \cite{Mannucci_2006}.
In FIRE-2, each WDSN produces $8.6 \times 10^{-3} \Msun$ of Mg and $0.74 \Msun$ of Fe, with IMF-averaged values from the W7 model in \cite{Iwamoto_1999}. Combined, these rate and yield models lead to $0.0011$ WDSN events per $\Msun$ per Hubble time ($13.7 \Gyr$), producing $6.6 \times 10^{-6} \Msun$ of Mg and $5.7 \times 10^{-4} \Msun$ per $\Msun$ per Hubble time.

For core-collapse supernovae (CCSN) rates, we use a step function based on energetics from \textsc{Starburst99} \citep{Leitherer_1999}, assuming each CCSN ejecta has $10^{51}$ erg.
`The specific event rate is $5.4 \times 10^{-4} \Msun^{-1} \, {\rm Myr}^{-1}$ across ages $3.4 - 10.37 \Myr$ and $2.5 \times 10^{-4} \Msun^{-1} \, {\rm Myr}^{-1}$ across ages $10.37 - 37.53 \Myr$. Each CCSN produces $9.9 \times 10^{-2} \Msun$ of Mg and $7.4 \times 10^{-2} \Msun$ of Fe, with IMF-averaged values from \cite{Nomoto_2006}. Combined, these rate and yield models lead to $0.11$ events per $\Msun$ per Hubble time ($13.7 \Gyr$), producing $1.0 \times 10^{-3} \Msun$ of Mg and $7.9 \times 10^{-4} \Msun$ of Fe per $\Msun$ per Hubble time.

Given the above, in FIRE-2, Mg is a nearly ``pure'' product of CCSN, so we use it as our fiducial $\alpha$ element.

\begin{figure*}[t]
\centering
\includegraphics[width = \linewidth]{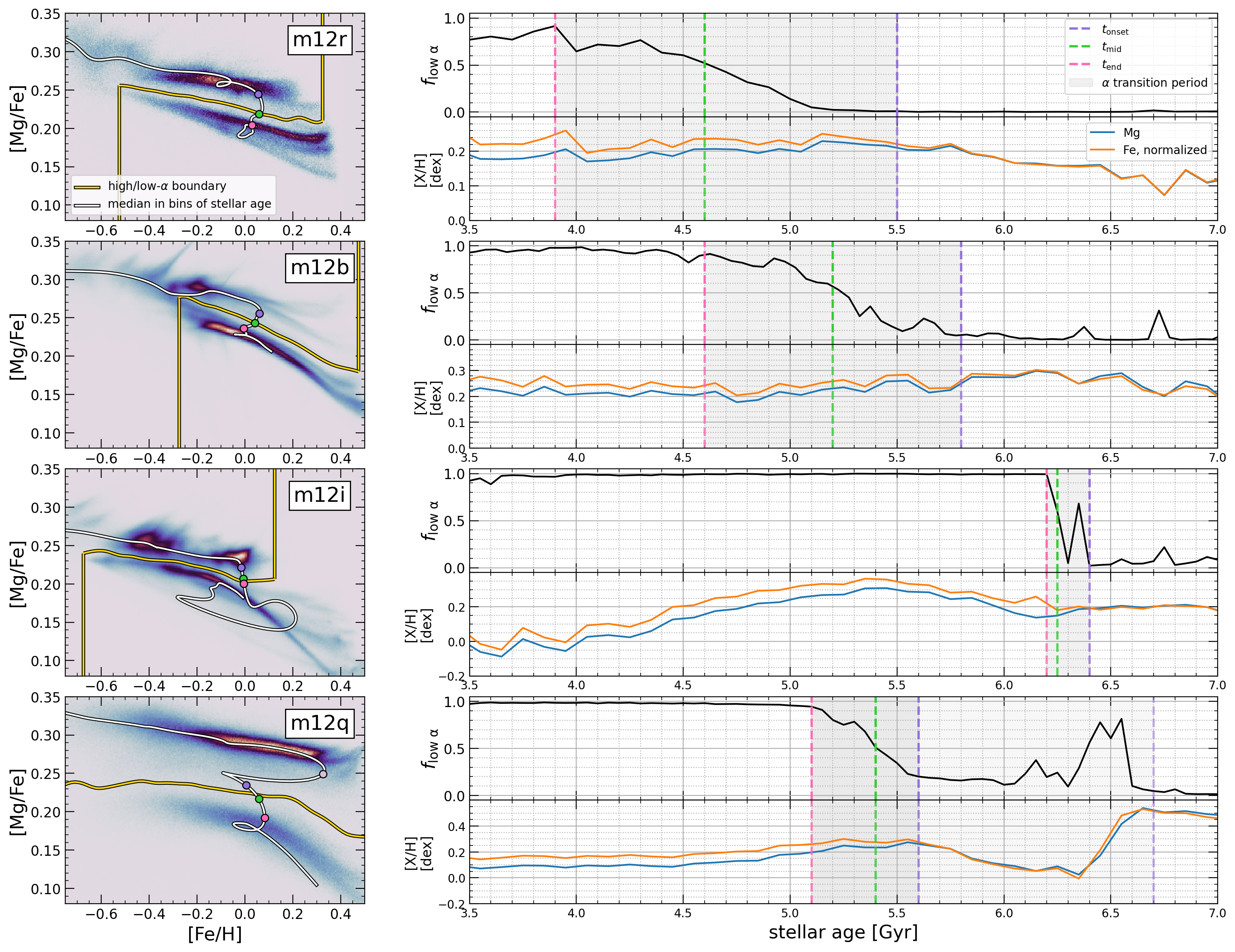}
\caption{
\textit{Left}: Histogram of [Mg/Fe] versus [Fe/H] today, as in Figure~\ref{fig:disks}.
Each white curve shows the path over time (beginning at the top left) of the median [Mg/Fe]-[Fe/H] of newly-formed stars across the galaxy.
Colored dots show the beginning ($t_{\rm onset}$), middle ($t_{\rm mid}$), and end ($t_{\rm end}$) of the high-to-low $\alpha$ transition, as defined in Section~\ref{sec:motiv} and listed in Table~\ref{tab:galaxies}.
\textit{Most galaxies underwent the $\alpha$ transition, evolving from high- to low-[Mg/Fe], at relatively fixed [Fe/H], following a downward ``turn'' in the evolutionary track.}
\textit{Right}: Top panels shows the low-$\alpha$ fraction, $f_{\rm low-\alpha} = M_{\rm star, low-\alpha} / M_{\rm star, tot}$, versus stellar age.
Vertical dashed lines indicate key times of the $\alpha$ transition: $t_{\rm onset}$, $t_{\rm mid}$, $t_{\rm end}$.
Shaded regions show the duration of the $\alpha$ transition, which ranges from $0.3 \Gyr$ (m12i) to $1.2 \Gyr$ (m12r).
We show 2 instances of $t_{\rm onset}$ for m12q (bottom), because $f_{\rm low-\alpha}$ rises but then falls to $\lesssim 0.2$ for nearly 1 Gyr before beginning to increase again.
Bottom panels show average [Mg/H] and [Fe/H] of stars at birth versus age.
We normalize [Fe/H] so [Mg/H] and [Fe/H] are equal $0.5 \Gyr$ before $t_{\rm onset}$ to compare their evolution and divergence.
During the $\alpha$ transition, [Fe/H] remained relatively fixed, via a near-balance between (strong) Fe enrichment and gas accretion (dilution), while [Mg/H] generally declined, because of less rapid Mg enrichment.
}
\label{fig:alphafrac}
\end{figure*}

\subsection{\texorpdfstring{Measuring a bimodality in $\alpha$-elements}{Measuring a bimodality in alpha-elements}}
\label{subsec:define}

We examine [Mg/Fe] versus [Fe/H] for 16 different MW-mass galaxies (from distinct initial conditions) in the FIRE-2 simulations.
Several FIRE-2 galaxies have resimulations with variations in resolution and/or the assumed UV background, which we also examine.
Our primary goal in this work is to explore the formation histories of galaxies that have a \textit{strong} bimodality in $\alpha$ elements, that is, a \textit{significant and well defined} gap (bimodality) in [Mg/Fe] over a wide range of [Fe/H], resulting in the presence of distinct high- and low-$\alpha$ populations.
We are motivated by understanding such patterns in the MW.
Focusing on strong bimodalities with distinct sequences also allows us to unambiguously identify stars as belonging to high- or low-$\alpha$ populations.
Some FIRE-2 galaxies exhibit weaker bimodalities in [Mg/Fe], in some cases across a narrower range of [Fe/H], while some FIRE-2 galaxies exhibit no clear bimodalities.
\cite{Parul_2025} presented [Mg/Fe] versus [Fe/H] for nearly all FIRE-2 MW-mass galaxies and explored some of these galaxies with a weaker and/or partial bimodality. Of these galaxies, they categorized 3 as ``non-bimodal'', 2 as ``weakly bimodal'', 5 as ``bimodal'', and 1 (m12b) as ``strongly bimodal''
We include that same version of m12b in this work.

\begin{figure*}[t]
\vspace{-3 mm}
\centering
\includegraphics[height=0.68\textheight,keepaspectratio]{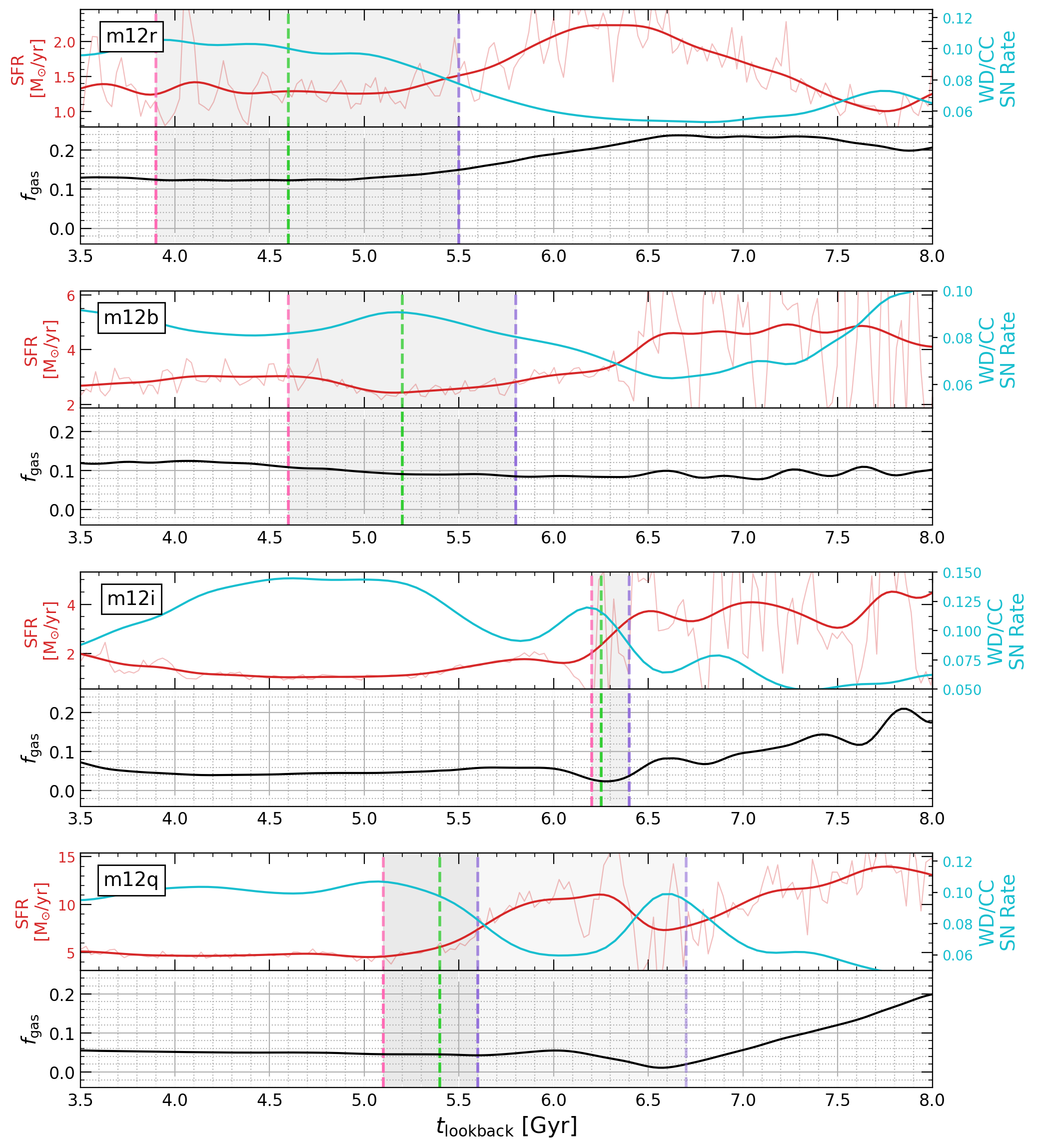}
\vspace{-4 mm}
\caption{
Key physical properties versus lookback time that drove the strong $\alpha$-element bimodality in these 4 galaxies.
Vertical dashed lines and gray shaded regions show the $\alpha$ transition period, as defined by $f_{\rm low-\alpha}$ in Figure~\ref{fig:alphafrac}.
\textit{Top rows}: Each red curve shows the star formation rate (SFR): the thick curve shows SFR smoothed over 50 Myr, highlighting a sharp drop (up to a twofold decrease) in SFR immediately before the $\alpha$ transition, while the thin curve shows SFR in bins of 10 Myr, which shows a transition from bursty to smooth star formation when the SFR declined for all cases except m12r.
Each blue curve shows the instantaneous ratio of the (Fe-producing) white-dwarf supernova (WDSN) rate to the ($\alpha$-producing) core-collapse supernova (CCSN) rate, which rose following the drop in SFR, typically approximately doubling.
In all 4 galaxies, the $\alpha$ transition occurred after or during a drop in SFR, corresponding to an increase in the WD/CC SN rate, and thus more enrichment in Fe than Mg.
\textit{Bottom rows}: Gas fraction, $f_{\rm gas} = M_{\rm gas} / (M_{\rm gas} + M_{\rm star})$, calculated within $R_{\rm star,95}$ and $Z_{\rm star,95}$ at that snapshot.
The $\alpha$ transition occurred after a $\approx 50\%$ decrease in $f_{\rm gas}$ for m12r, m12i, and m12q.
In all 4 cases, $f_{\rm gas}$ remained nearly constant across the transition period, despite ongoing star formation, indicating ongoing gas accretion and dilution.
}
\label{fig:sfr}
\end{figure*}

We identified 4 galaxies that meet our criteria of a ``strong'' bimodality: m12r, m12b, m12i, and m12q. (We note that this is a different variation of m12i than the non-bimodal version discussed in \cite{Parul_2025}.)
Table~\ref{tab:galaxies} lists their properties.
\cite{Parul_2025} showed [Mg/Fe] versus [Fe/H] for m12b and the fiducial versions of m12i and m12r.
However, they did not examine m12q, which we present here for the first time.
We find that a lower-resolution version of m12r, and a version of m12i that uses the updated UV background from \cite{Faucher_2020}, both produce a stronger bimodality than their fiducial FIRE-2 versions, so we present them here.
We do not think that the resolution or the UV background directly affects the presence of a strong bimodality, but rather, we suspect that the presence of a (strong) bimodality is a result (at least in part) of stochastic effects, including the details of the formation/merger history, and the exact timing of changes to star formation and its relation to gas content.
Indeed, recent work has shown that resimulating the same galaxy with the same resolution and physics can lead to significantly different properties, given the chaotic nature of galaxy formation \citep{Keller_2019, Genel_2019, Borrow_2023}.
In future work, we will investigate multiple resimulations of individual galaxies (using the same initial conditions) to understand this more fully.
In this work, we focus on understanding why these versions of these 4 galaxies became strongly bimodal.

For each MW-mass galaxy, we measure stars today within galactocentric $R < 15 \kpc$ and $|Z| < 3 \kpc$.
We define a boundary that delineates stars in ``high-$\alpha$'' and ``low-$\alpha$'' populations. We define this boundary by first separating stars into 0.05 dex bins of [Fe/H].
For bins that have a well-defined minimum between high- and low-$\alpha$ populations, we compute the minimum of the distribution of [Mg/Fe].
We then apply a cubic spline interpolation in the [Mg/Fe] break (minimum) across the range of [Fe/H] to form a smooth boundary.
Such a well-defined bimodality in [Mg/Fe] does not always extend across all bins of [Fe/H]: depending on the galaxy, 5 - 30\% of stars lie outside the range of [Fe/H] with a well-defined bimodality in [Mg/Fe].
For these stars, if they are at lower [Fe/H] (and typically higher [Mg/Fe]), we associate them with the high-$\alpha$ population, and if they are at higher [Fe/H] (and typically lower [Mg/Fe]), we associate them with the low-$\alpha$ population.

Within the range of [Fe/H] with a well-defined bimodality, we measure the ``strength'' of the bimodality via $\Delta \rm [Mg/Fe]$, the ``gap width'' in [Mg/Fe] at fixed [Fe/H] (using 0.2 dex bins of [Fe/H]) between the peaks (modes) of the high- and low-$\alpha$ populations.
Because this gap width can vary with [Fe/H], we in Table~\ref{tab:galaxies} we report the average across the range of [Fe/H] with a well-defined bimodality, as the yellow curves in Figure~\ref{fig:disks} indicate.

\section{Results}
\label{sec:results}

\subsection{Properties today}
\label{sec:motiv}

The physical processes that drive changes in stellar abundances in Mg and Fe are primarily CCSN and WDSN, respectively.
CCSN occur promptly after star formation, within $\approx 40 \Myr$, and produce $\alpha$-elements like Mg and also Fe.
WDSN are more delayed and produce primarily Fe.
Therefore, as a galaxy evolves and enriches in [Fe/H], it generally should decrease in [Mg/Fe].

\begin{figure*}[t]
\centering
\includegraphics[width = \linewidth]{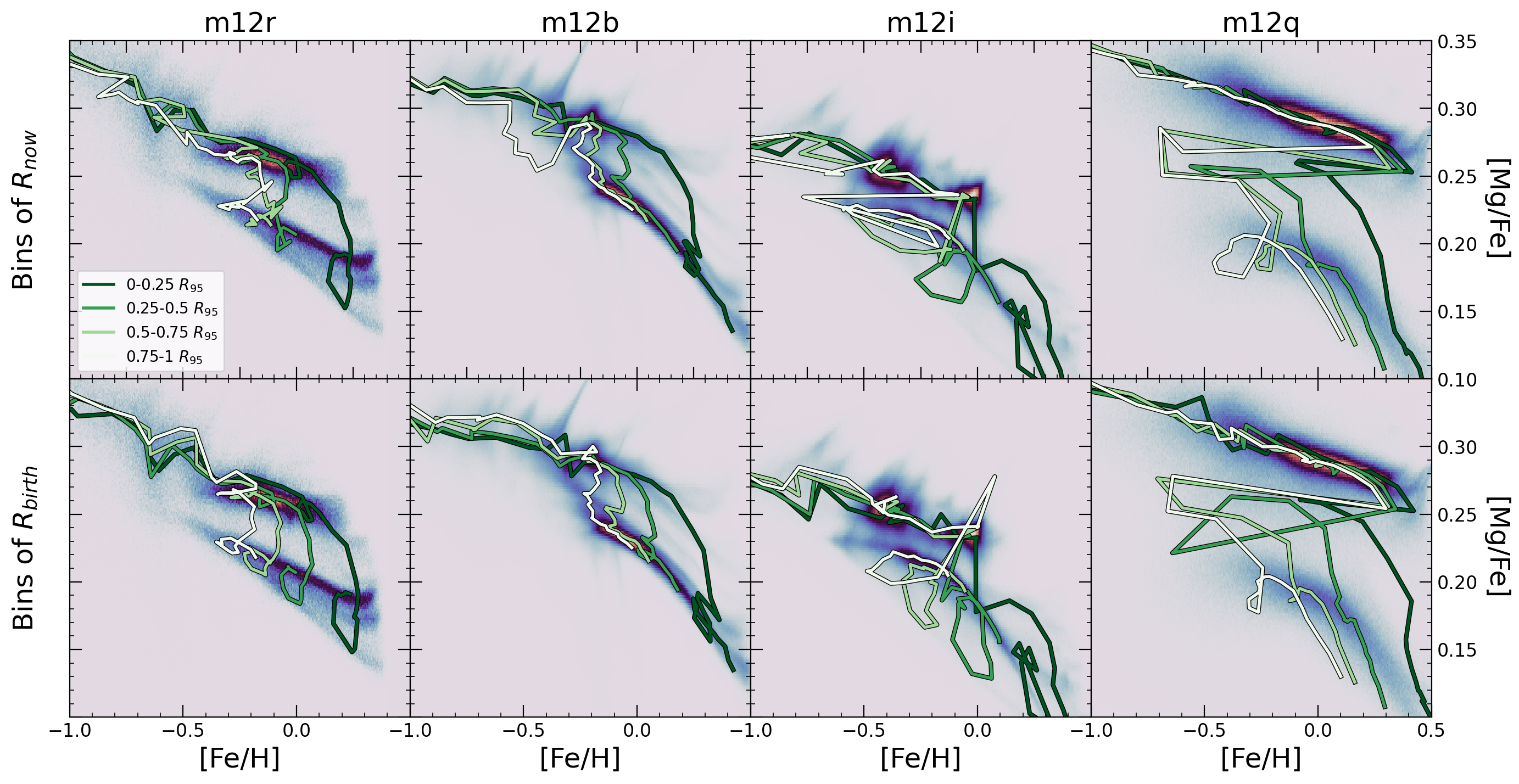}
\caption{
Paths through [Mg/Fe]-[Fe/H], as in Figure~\ref{fig:alphafrac}, now split into bins of galactocentric radii, for both $R_{\rm star,now}$ (top) and $R_{\rm star,now}$ (bottom), in units of $R_{\rm star,95}$.
Both show similar trends, further highlighting that the key patterns of the $\alpha$ bimodality were present for stars at birth and were not significantly affected by radial redistribution of stars after birth.
The erratic behavior in m12i arises from bins of low star formation.
Overall, stars at different $R$ had similar abundances during the high-$\alpha$ period, but stars in different $R$ split apart around the time of the transition to the low-$\alpha$ phase, because this coincided with the formation of a radial gradient in metallicity.
Even at fixed $R$, stars evolved from high- to low-[Mg/Fe] during the $\alpha$ transition period at relatively fixed [Fe/H].
}
\label{fig:paths}
\end{figure*}

\begin{figure}
\centering
\includegraphics[height=0.88\textheight,keepaspectratio]{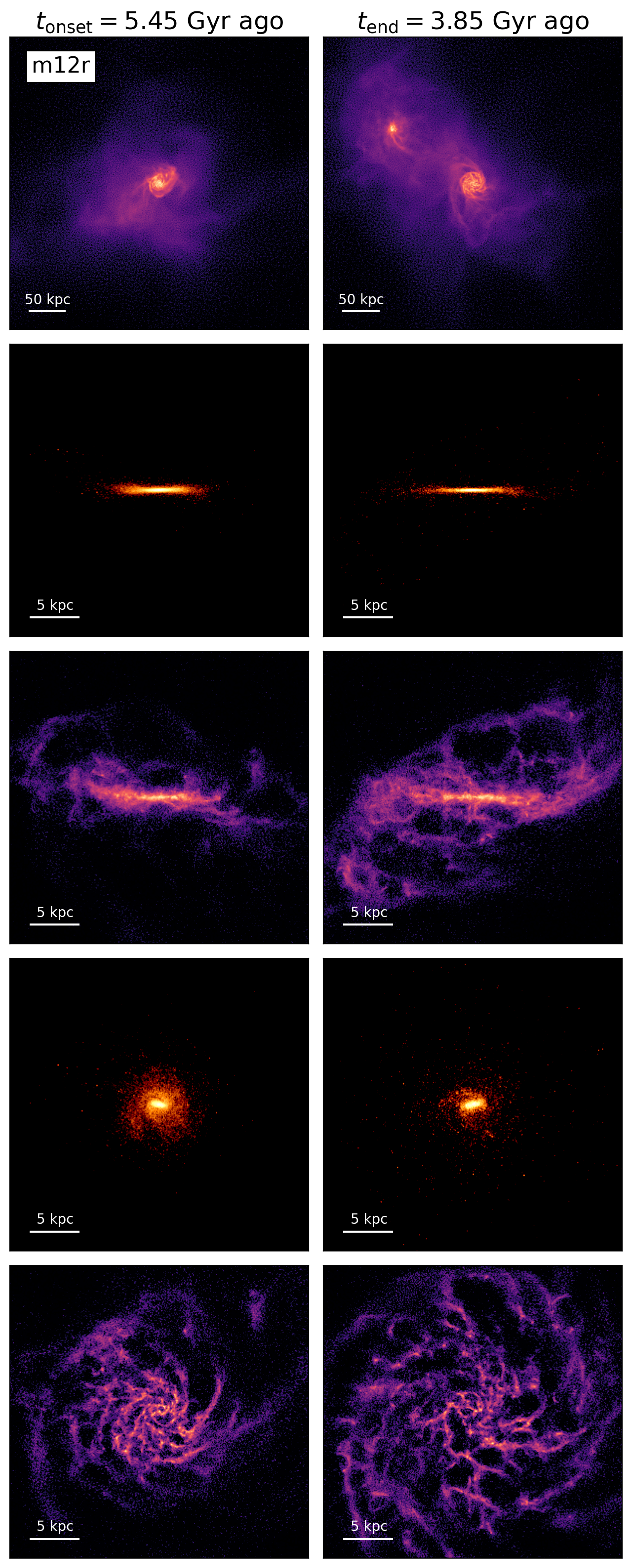}
\caption{
\textbf{m12r}: images of gas (top, third, fifth row) and recently formed stars (second, fourth row) at the beginning (left) and end (right) of the $\alpha$ transition period.
Top row is zoomed out, while all other panels zoom in on the disk, both edge-on and face-on.
See Section~\ref{sec:indiv} for discussion.
}
\label{fig:m12r}
\end{figure}

Figure~\ref{fig:disks} (left) shows histograms of [Mg/Fe] versus [Fe/H] today for our 4 strongly bimodal galaxies.
As expected, the oldest stars populate the upper left, enriched by CCSN but not yet significantly enriched in the additional Fe from WDSN, while increasingly younger stars generally increase in [Fe/H] and decrease in [Mg/Fe], as Fe production by WDSN becomes more significant.
These abundances are systematically high in [Mg/Fe] compared to the MW (by $\sim0.1-0.2$ dex), likely because of uncertainties in our assumed supernova rate and yield models, especially the WDSN DTD, which may underproduce Fe \citep{Gandhi_2022}.
However, our focus in this work is on relative abundances, their changes over time, and features in [Mg/Fe]-[Fe/H] space, rather than absolute values of these abundances.
In future work, we will explore the extent to which changes to our assumed enrichment model change these features in [Mg/Fe]-[Fe/H].

The yellow curve in Figure~\ref{fig:disks} shows the delineation between well-defined high- and low-$\alpha$ populations, as in Section~\ref{subsec:define}.
The ``gap width'' between the peaks (modes) of the high- and low-$\alpha$ populations, $\Delta[{\rm Mg}/{\rm Fe}]$, ranges from 0.03 dex for m11i to 0.11 dex for m12q (Table~\ref{tab:galaxies}).

Figure~\ref{fig:disks} (right) shows images of the high- and low-$\alpha$ populations for each galaxy today, in a side-on view of the disk.
In all 4 cases, high-$\alpha$ stars correspond to an older, radially compact ``thick'' disk.
Across these 4 galaxies, high-$\alpha$ stars have average age $\langle t_{\rm birth} \rangle = 8.5 \Gyr$ ago, radius $\langle R_{\rm star,95} \rangle = 7.5$ kpc, and height $\langle Z_{\rm star,95} \rangle = 1.6$ kpc, where $R_{\rm star,95}$ and $Z_{\rm star,95}$ are the $R$ and $Z$ values containing 95\% of the galaxy's stellar mass.
By contrast, low-$\alpha$ stars correspond to a younger, radially extended ``thin'' disk, with average age $\langle t_{\rm birth} \rangle = 3.1 \Gyr$ ago, $\langle R_{\rm star,95} \rangle = 10.3 \kpc$, and $\langle Z_{\rm star, 95} \rangle = 1.0 \kpc$.
These relations between elemental abundance, age, radial extent, and thickness are qualitatively consistent with observations of MW  \citep[for example,][]{Fuhrmann_1998, Bensby_2003, Reddy_2003, Bensby_2005, Bovy_2012a, Bensby_2014, Anders_2014, Hayden_2015, Masseron_2015, Wojno_2016, Feuillet_2019, Vincenzo_2021, Horta_2023, Ratcliffe_2023, Imig_2023, Khoperskov_2024, Borbolato_2025}.
This also reinforces results from \cite{Graf_2024}, who quantified how these
FIRE-2 galaxies exhibit inside-out radial growth and upside-down disk settling: early on, stars form in a compact, kinematically hot, metal-poor gas disk but as ISM turbulence and radial gas flows decline and the disk becomes more rotationally supported, star formation shifts to larger radii and lower scale heights, producing a younger, more metal-rich, low-$\alpha$ thin disk surrounding the older, high-$\alpha$ thick disk.
Essentially all FIRE-2 MW-mass galaxies exhibit similar relations between stellar age, radius, thickness, and elemental abundances, even if they are not bimodal \citep{Yu_2023, McCluskey_2024}, but the presence of a well-defined gap between high- and low-$\alpha$ stars in these 4 galaxies allows us to split them by elemental abundances, as many works do for the MW.

One notable difference from the MW is that the low-$\alpha$ population appears in these galaxies significantly later than it does in the MW ($\approx 8 - 10 \Gyr$ ago) \citep{SilvaAguirre_2018, Khoperskov_2024, Thulasidharan_2024}.
As a result, these FIRE-2 galaxies also have more radially-extended high-$\alpha$ populations.
Furthermore, all 4 FIRE-2 galaxies retain distinct high- and low-$\alpha$ tracks even at high [Fe/H].
By contrast, many observational analyses suggest that the tracks in the MW converge at [Fe/H] $\gtrsim 0$, although the nature and degree of this convergence remain debated \citep{Conroy_2022, Hayes_2022, Buder_2025}.


Figure~\ref{fig:hayden} (left) shows [Mg/Fe] versus [Fe/H] histograms for each galaxy today, separated into spatially-defined slices of galactocentric radial radius (in units of $R_{\rm star,95}$) and vertical height ($|Z|$, in kpc). For all 4 galaxies, the high-$\alpha$ population dominates at high $|Z|$ and small $R$, while the low~$\alpha$ population dominates at low $|Z|$ and large $R$, similar to the MW \citep{Anders_2014, Nidever_2014, Hayden_2015, Vincenzo_2021}. However, unlike the MW, these galaxies have more high-$\alpha$ stars at low $|Z|$ and small $R$, likely because these galaxies underwent the transition to low~$\alpha$ later than the MW.
m12q and m12r also have some high-$\alpha$ stars at larger $R$, as also visible in Figure~\ref{fig:disks}, because the radial extent of their high- and low-$\alpha$ populations are more similar, again likely because of their relatively later transition to low~$\alpha$.

Figure~\ref{fig:hayden} (right) shows the same histograms, but for stars at their birth coordinates.
Here we scale $R_{\rm birth}$ to $R_{\rm star,95}$ today (not at birth), such that the scaling of the x-axis remains the same.
These results are overall similar to those for stars based on their coordinates today.
The main difference is that, at birth, fewer stars were at high $|Z|$, especially at smaller $R$, consistent with the dynamical heating of stars over time \citep{Hayden_2015, Schonrich_2017, Sharma_2021, McCluskey_2024}, but this does not qualitatively change the spatial distributions of high- and low-$\alpha$ populations.
Thus, at least for these FIRE-2 galaxies, the existence of a bimodality or the locations of high- and low-$\alpha$ populations today are not primarily a result of radial redistribution, as some works argue \citep{Schonrich_2009, Sharma_2021}.

\subsection{Common trends in formation history}
\label{sec:trends}

We next examine the formation histories of these 4 galaxies, before and during their transition from high- to low-$\alpha$, to understand the origin of such strong bimodalities.
Here we focus on trends that these 4 galaxies have in common, while in Section~\ref{sec:indiv} we examine the specific history of each galaxy.

For each galaxy, we define 3 key lookback times, each relating to the fraction of low-$\alpha$ stars born at a given time, $f_{\rm low-\alpha} = M_{\rm star, low-\alpha} / M_{\rm star, tot}$.
First, $t_{\rm onset}$, when $f_{\rm low-\alpha}$ began to increase significantly ($\gtrsim 0.05$); $t_{\rm mid}$, when $f_{\rm low-\alpha} = 0.5$; and $t_{\rm end}$, when $f_{\rm low-\alpha}$ stabilized near 1.

Figure~\ref{fig:alphafrac} (right, black curves) shows $f_{\rm low-\alpha}$ over time for each galaxy.
The vertical lines indicate $t_{\rm onset}$, $t_{\rm mid}$, and $t_{\rm end}$, and the entire $\alpha$ transition period is shaded gray.
The duration of this $\alpha$ transition ranges from 0.28 to 1.6 Gyr, with a mean of 0.86 Gyr (see Table~\ref{tab:galaxies}).
The MW likely had an earlier low-$\alpha$ onset than these 4 galaxies, with most estimates being $7 - 9 \Gyr$ ago \citep{Haywood_2013, Haywood_2016, Snaith_2015, Feuillet_2019}.
This is likely why the MW has a larger low-$\alpha$ fraction in stars today ($f_{\rm low-\alpha} \approx 0.8$) than these 4 galaxies, which have $f_{\rm low-\alpha} = 0.35 - 0.46$.

Figure~\ref{fig:alphafrac} (right, bottom subpanels) shows the average [Mg/H] and [Fe/H] of all stars versus age.
We normalize these such that [Mg/H] = [Fe/H] 0.5 Gyr before $t_{\rm onset}$, to show their relative evolution during the $\alpha$ transition period.
During the transition of all 4 galaxies, [Fe/H] remained relatively unchanged over time.
From $t_{\rm onset}$ to $t_{\rm end}$, [Fe/H] changed by as little as 0.002 (m12i) and as much as 0.045 (m12r, for which it increased).
This indicates that Fe enrichment was offset by gas accretion/dilution.
By contrast, [Mg/H] typically decreased during this period, indicating that Mg enrichment was less rapid than that of Fe and could not keep up with gas accretion/dilution.
From $t_{\rm onset}$ to $t_{\rm end}$, [Mg/H] reduced by as little as 0.002 (m12r, which increased in [Fe/H]) and as much as 0.056 (m12b).

\begin{figure}
\centering
\includegraphics[height=0.88\textheight,keepaspectratio]{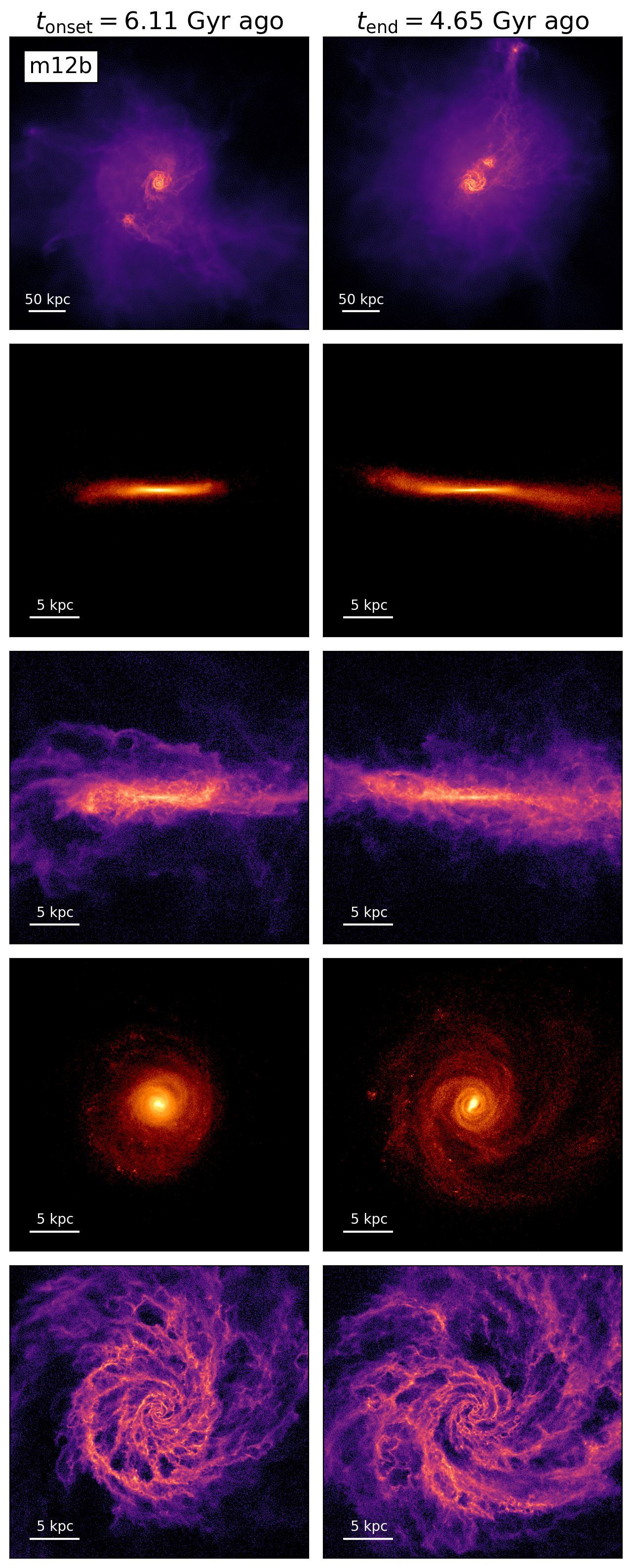}
\caption{
\textbf{m12b}: images of gas (top, third, fifth row) and recently formed stars (second, fourth row) at the beginning (left) and end (right) of the $\alpha$ transition period.
Top row is zoomed out, while all other panels zoom in on the disk, both edge-on and face-on.
See Section~\ref{sec:indiv} for discussion.
}
\label{fig:m12b}
\end{figure}

\begin{figure}
\centering
\includegraphics[height=0.88\textheight,keepaspectratio]{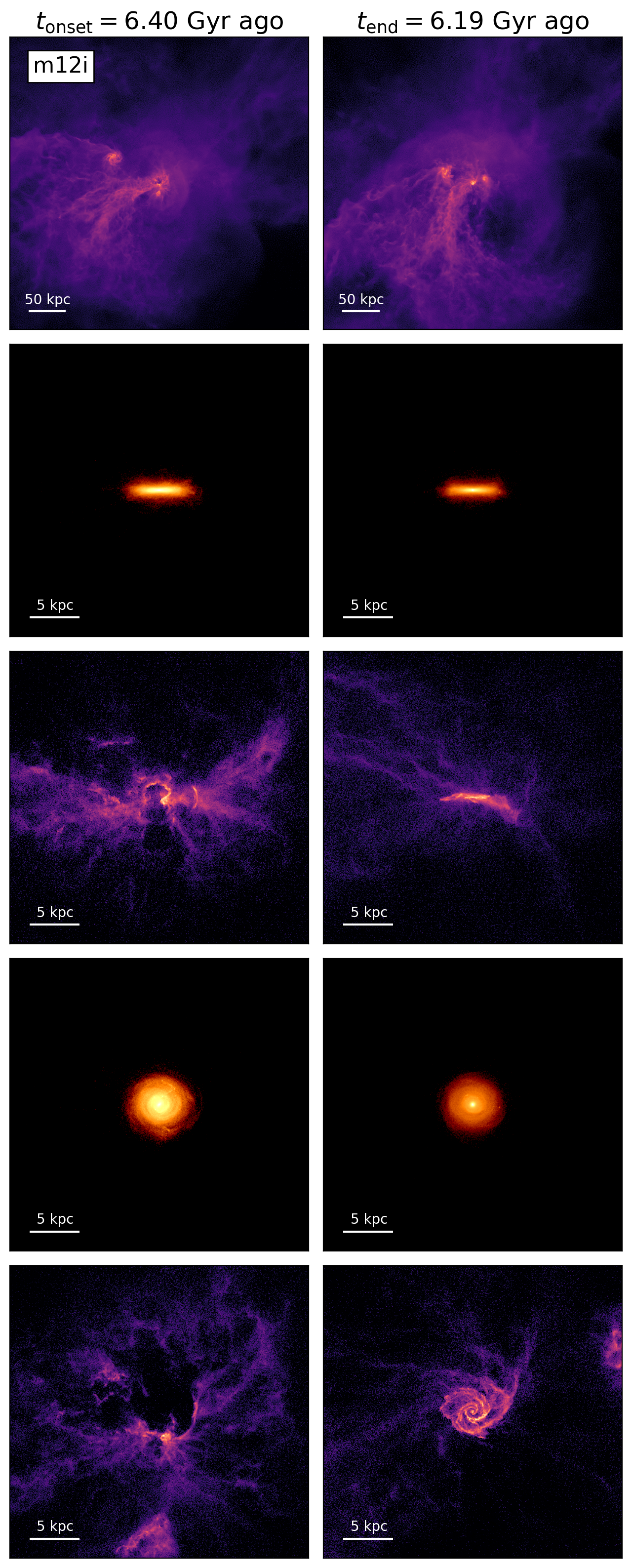}
\caption{
\textbf{m12i}: images of gas (top, third, fifth row) and recently formed stars (second, fourth row) at the beginning (left) and end (right) of the $\alpha$ transition period.
Top row is zoomed out, while all other panels zoom in on the disk, both edge-on and face-on.
See Section~\ref{sec:indiv} for discussion.
}
\label{fig:m12i}
\end{figure}

Figure~\ref{fig:alphafrac} (left) shows the path that each galaxy took through [Mg/H] versus [Fe/H]. Specifically, including all stars within the galaxy, we divide stars into 50 Myr bins of stellar age and compute the median [Mg/Fe] and [Fe/H] at each age.
We then plot this track using a smoothing window of 0.25 Gyr.
Each path begins at the upper left, and colored dots show $t_{\rm onset}$, $t_{\rm mid}$, and $t_{\rm end}$ of the $\alpha$ transition period, as in the right panels.

During the $\alpha$ transition period, [Mg/Fe] decreased at roughly fixed [Fe/H] in \textbf{all} galaxies. For 3 of these galaxies (m12r, m12b, m12i), the change in [Fe/H] from $t_{\rm onset}$ to $t_{\rm end}$ less than 0.05 dex. The fourth galaxy, m12q, underwent more complex behavior: after $f_{\rm low-\alpha}$ spiked at a lookback time of $\approx 6.8$ Gyr, m12q returned to forming stars predominantly at high-$\alpha$ ($f_{\rm low-\alpha} \lesssim 0.2$) for the following $\approx 1 \Gyr$. Thus, we consider $t_{\rm onset}$ for m12q to be when $f_{\rm low-\alpha}$ began to rise permanently, for the second time, $\approx 5.6 \Gyr$ ago.
We discuss this in Section~\ref{sec:indiv}.
Among these 4 galaxies, shorter $\alpha$ transition periods lead to smaller gap widths between the bimodal tracks (defined in \ref{sec:motiv}).

To explain this decrease in [Mg/Fe] at fixed [Fe/H], Figure~\ref{fig:sfr} shows the evolution of 3 key parameters relevant to elemental abundances: (1) the star formation rate (SFR), (2) the ratio of WDSN to CCSN, which is set by the SFR history, and (3) the gas fraction $f_{\rm gas} = M_{\rm gas} / (M_{\rm gas} + M_{\rm star})$. Again, the shaded region shows the $\alpha$ transition period.

Figure~\ref{fig:sfr} shows the strongest unifying features among these 4 strongly bimodal galaxies.
Specifically, (1) the $\alpha$ transition occurred after a significant and rapid decrease in SFR, typically by a factor of $\approx 2$ over $\lesssim 500 \Myr$, and (2) during a period of low $f_{\rm gas}$, in these cases, $0.02 - 0.14$.
Furthermore, 3 galaxies (m12r, m12i, m12q) experienced a significant decline in $f_{\rm gas}$ right before the transition, commensurate with their decline in SFR.
The other, m12b, underwent a substantial decline in $f_{\rm gas}$ well before the $\alpha$ transition (from 0.45 to 0.1 across $10$ to $8 \Gyr$ ago).

Importantly, as Figure~\ref{fig:sfr} shows via blue lines, after the SFR decreased, a higher fraction of Fe-producing WDSN occurred relative to $\alpha$-producing CCSN, given their differing DTDs.
Again, this boost in relative SN rates is substantial, typically a factor of $\approx 2$ increase over $0.3 - 1 \Gyr$.
This led to a rapid decrease in [Mg/Fe], which was able to be rapid because of the low $f_{\rm gas}$.
In other words, \textit{the small gas reservoir (leading to less dilution) allowed the ISM to enrich rapidly enough to create a gap in [Mg/Fe], even absent a gap in SFR.}

A number of works using analytic models or simulations argued for the need for a complete pause in SFR between high- and low-$\alpha$ formation to produce a bimodality \citep[for example][]{Lian_2020, Spitoni_2024, Beane_2025}. However, we find no instances of a complete pause in SFR among our FIRE sample; only a significant, permanent decrease. As Figure~\ref{fig:sfr} shows, the SFR declines by at most a factor of $\sim 2$ across the $\alpha$ transition period, but there is never a true quiescent phase where the SFR falls to zero.
\textit{Therefore, these FIRE simulations demonstrate that a complete pause in SFR is not necessary to produce a strong bimodality.}

A related unifying feature of these 4 galaxies is that, before the $\alpha$ transition period, stellar populations evolved through [Mg/Fe]-[Fe/H] space as expected, that is, diagonally downward from top left to bottom right, as they enriched in Fe, and in Fe more rapidly than in Mg.
However, during the transition to low-$\alpha$, this evolution changed ``direction'' in this space, evolving typically straight downward, decreasing in [Mg/Fe] at roughly fixed [Fe/H].
This is because the (relatively) stronger Fe enrichment nearly offset the accretion/dilution from new gas, but the weaker Mg enrichment did not.
We hypothesize that this is another key feature that promotes a strong bimodality.

Figure~\ref{fig:paths} shows this evolution split into bins of galactocentric radii, both as measured for stars today (top) and at birth (bottom).
Importantly, the general trends are similar for binning by $R_{\rm birth}$ and $R_{\rm now}$, so radial redistribution did not play a significant role in the formation of an $\alpha$ bimodality, and it did not play a dominant role in its spatial pattern in stars today.
In general, tracks at fixed $R$ show similar behavior to the galaxy overall, in particular, evolution ``downward'' in [Mg/Fe] at relatively fixed [Fe/H].
Furthermore, the $\alpha$ transition period generally coincided with the onset of a radial gradient in [Fe/H], with stars at a given time at smaller $R$ having higher [Fe/H].
In other words, high-$\alpha$ stars formed when the galaxy was well-mixed radially, with stars of a given age forming with relatively similar elemental abundances.
By contrast, low-$\alpha$ stars formed when the galaxy sustained a wide range of [Fe/H] at a given age.
Because the onset of a radial gradient occurs during the $\alpha$ transition period for all 4 galaxies, this may be a key condition to produce a strong bimodality.
However, \citep{Bellardini_2022} showed that essentially all MW-mass FIRE-2 galaxies form negative gradients in metallicity, so this may be a necessary but not sufficient condition.

Figure~\ref{fig:sfr} also shows that in 3 galaxies (m12b, m12i, m12q), SFR was also burstier (more time variable) before $t_{\rm onset}$.
While the SFR may appear bursty even at later times in m12r,  Figure~\ref{fig:paths} shows a narrower dynamic range in SFR for m12r, in part because m12r underwent a transition from highly bursty to relatively smoother SFR earlier than Figure~\ref{fig:paths} shows.
Therefore, the level of burstiness in m12r is comparable to the other galaxies at late times.
The similarity of m12b, m12i, and m12q may suggest that the transition from bursty to smooth SFR helps promote the onset of low-$\alpha$ stars.
Indeed, \citet{Parul_2025} argued for such a correlation across a larger set of FIRE-2 galaxies.
However, we argue that the most direct cause is the longer-term ratio of WDSN/CCSN rates, which determines the relative enrichment of Fe versus Mg.
That said, the decline in (average) SFR generally coincides with a decline in burstiness (except for m12r), and the decline in burstiness also helps promote stronger metallicity radial gradients by reducing the amount of radial mixing in gas  \citep{Graf_2024}, so these are linked phenomena.
In future work, we will attempt to disambiguate their effects on abundance patterns more directly.

All 4 galaxies had an established (thick) disk before $t_{\rm onset}$.
Table~\ref{tab:galaxies} lists the lookback time that each galaxy first formed a long-lived ``disk'', using the criteria from \cite{McCluskey_2024}: $v_{\phi} / \sigma_{\rm v,3D} > 1$, measured for stars at birth, which was typically $\approx 8 \Gyr$ ago, at the right edge of Figure~\ref{fig:sfr}.
Two galaxies (m12i and m12q) experienced temporary disruptions of their disk during the $\alpha$ transition period, visible as temporary spikes in $f_{\rm low-\alpha}$, which we discuss in the next subsection.
We emphasize that the disk was not dramatically thinner immediately as the low-$\alpha$ stars started to form; rather, the disk continued to settle throughout the low-$\alpha$ period. The average $v_{\phi} / \sigma_{\rm v,3D}$ across all four galaxies was 1.6 at $t_{\textrm{onset}}$ and 1.98 at $t_{\textrm{end}}$, with an average difference of 0.38. m12i underwent an extreme, short-lived disk disruption during $t_{\textrm{onset}}$; if we exclude m12i, this difference drops to just 0.12.

In summary, in all 4 strongly bimodal galaxies, the onset of the $\alpha$ transition occurred after a rapid drop ($\approx 2 \times$) in SFR and a corresponding rise ($\approx 2 \times$) in the ratio of WDSN/CCSN rates.
This occurred during a period of low $f_{\rm gas} = 0.02 - 0.14$, and in 3 of the 4 cases $f_{\rm gas}$ rapidly declined before the transition.
Combined, this allowed for a rapid decline in [Mg/Fe], which also occurred at roughly fixed [Fe/H], from a balance between Fe enrichment and gas accretion/dilution.
\textit{However, as we describe below, there is no commonality among the merger histories of these galaxies.}

\subsection{Histories of individual galaxies}
\label{sec:indiv}

We now discuss each galaxy individually, to highlight differences between them, including merger events and accretion that occurred before and during the $\alpha$ transition period.
Overall, we find no particular unifying merger history tied to the $\alpha$ bimodality.
In Figures~\ref{fig:m12r}~-~\ref{fig:m12q}, we show images of gas and stars in each galaxy at the beginning ($t_{\rm onset}$, left) and end ($t_{\rm end}$, right) of the $\alpha$ transition period.
In each figure, the top shows a zoomed-out view (400 kpc across) of total gas density.
The remaining rows show zoomed-in views (25 kpc across), both side-on (rows 2 and 3) and face-on (rows 4 and 5) with respect to the disk, of both gas and recently formed stars (age $< 0.5 \Gyr$).

\subsubsection{m12r}

Among our 4 strongly bimodal galaxies, m12r has the lowest stellar mass ($3.9 \times 10^{10} \Msun$ today), the lowest SFR, and both the latest and longest $\alpha$ transition, occurring $5.5 - 3.9 \Gyr$ ago.
m12r underwent no merger or direct interaction with a major satellite galaxy during this period.
$\approx 8 \Gyr$ ago, a massive satellite galaxy reached its first close approach and began an extended ($\approx 2 \Gyr$) close-range interaction with the galaxy, triggering a sharp increase in SFR that reached a peak at $\approx 6.5 \Gyr$ ago.
Gas deposition from this satellite is visible as a bump in $f_{\rm gas}$ in Figure~\ref{fig:sfr}.
In this galaxy, the decrease in SFR and corresponding increase in the ratio of WDSN/CCSN rates occurred as this accretion came to an end.
Figure~\ref{fig:m12r} shows that m12r's disk morphology is similar at the beginning and end of the $\alpha$ transition period.
The satellite galaxy visible in the top right panel is on its first infall and unrelated to the $\alpha$ transition.

In the other 3 galaxies, the $\alpha$ transition period occurred not just after a drop in (time-averaged) SFR, but after a transition from bursty to smooth SFR.
While m12r may appear to be highly bursty across the time range in Figure~\ref{fig:sfr}, the dynamic range of its SFR in Figure~\ref{fig:sfr} is much smaller than the other galaxies; in fact, m12r transitioned from bursty to smooth $> 8 \Gyr$ ago, not visible in Figure~\ref{fig:sfr}.
Thus, we conclude that in the case of m12r, it is the overall decrease in (time-averaged) SFR, not the decrease in burstiness, that is relevant to the formation of a bimodality.

\subsubsection{m12b}

m12b experienced a longer $\alpha$ transition period, with a duration of 1.35 Gyr. This galaxy is an outlier in several ways, most noticeably in that no significant events occurred near the $\alpha$ transition period, in terms of mergers or changes in $f_{\rm gas}$. Rather, it underwent relatively smooth gas accretion/dilution that balanced Fe production, such that [Mg/Fe] decreased at fixed [Fe/H]. 
Its lack of a major merger at that time may relate to its lack of a Fe-poor low-$\alpha$ population, compared to the other 3 galaxies.
Similar to m12r, Figure~\ref{fig:m12b} shows that m12b maintained a stable disk throughout the $\alpha$ transition period, though it grew radially somewhat.
We note that the satellite visible in the first row of Figure~\ref{fig:m12b} is $\sim50$ kpc away in the $z$-direction; we show this angle to give a top-down view of the disk.

\subsubsection{m12i}

m12i experienced the earliest and shortest $\alpha$ transition, which lasted $\lesssim 300 \Myr$.
As a result, it has the smallest gap width (0.03 dex) in [Mg/Fe] between high- and low-$\alpha$ populations.
m12i first met the criterion for disk onset (from \cite{McCluskey_2024} $8.4 \Gyr$ ago, though the gas disk underwent a cycle of feedback-induced blowouts over the next 2 Gyr.
The $\alpha$ transition occurred $6.4 \Gyr$ ago, during the last of these blowouts, visible in Figure~\ref{fig:m12i} (3rd and 5th rows), during which nearly all gas was driven out of the galaxy ($f_{\rm gas} \lesssim 0.02$, see Figure~\ref{fig:sfr}).
Stars formed in bursts during the following $0.1 \Gyr$, with the SFR oscillating between 0 and as high as 4 $M_{\odot} {\rm yr}^{-1}$ or more. After this blowout, the gas disk reformed permanently, visible $6.2 \Gyr$ ago in Figure~\ref{fig:m12i} (5th row).

Intriguingly, m12i also exhibits multimodality in its high-$\alpha$ population.
This likely resulted from extreme bursty star formation in its early history, as \cite{Patel_2022} and \cite{Parul_2025} explored for lower-mass and MW-mass galaxies in FIRE-2 simulations, respectively.

\subsubsection{m12q}

\begin{figure}
\centering
\includegraphics[width = \linewidth]{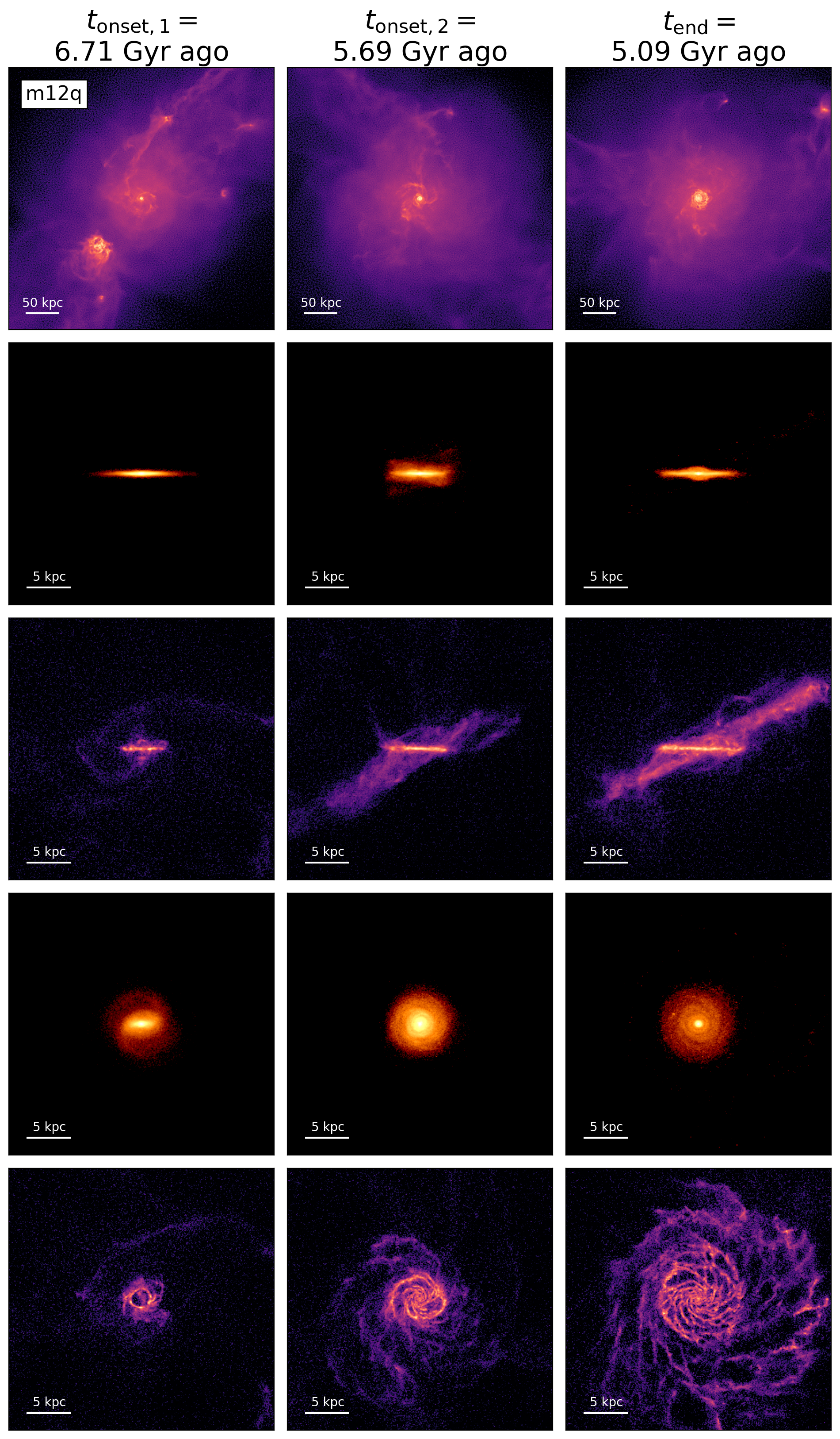}
\caption{
\textbf{m12q}: images of gas (top, third, fifth row) and recently formed stars (second, fourth row).
Top row is zoomed out, while all other panels zoom in on the disk, aligned both edge-on and face-on.
See Section~\ref{sec:indiv} for discussion.
\textit{Left columns}: The ``first start'' of its $\alpha$ transition (corresponding to the light gray shaded region in Figures~\ref{fig:alphafrac} and \ref{fig:sfr}).
The transition to low-$\alpha$ began while a massive ($M_{\rm star,peak} \approx 4 \times 10^{9} \Msun$), gas-rich ($M_{\rm gas} \approx 1.4 \times 10^{10} \Msun$, approximately equal to $M_{\rm gas}$ of the main galaxy) galaxy merged with it (visible in the lower left of the top panel).
After this merger, $f_{\rm low-\alpha}$ briefly spiked but then dropped to $\approx 0.2$, where it stayed roughly constant for $\approx 1 \Gyr$, before increasing again nearly 1 Gyr later.
During this, the gas disk of m12q briefly disappeared.
\textit{Center columns}: the ``main start'' of its $\alpha$ transition.
By this time, the gas disk reformed, and SFR began to drop again.
\textit{Right columns}: the end of its $\alpha$ transition, showing the deposition of gas from the satellite. This continual gas accretion drove this galaxy's post-merger drop in [Mg/Fe] at relatively fixed [Fe/H].
}
\label{fig:m12q}
\end{figure}

\begin{figure}[t]
\centering
\includegraphics[width = \linewidth]{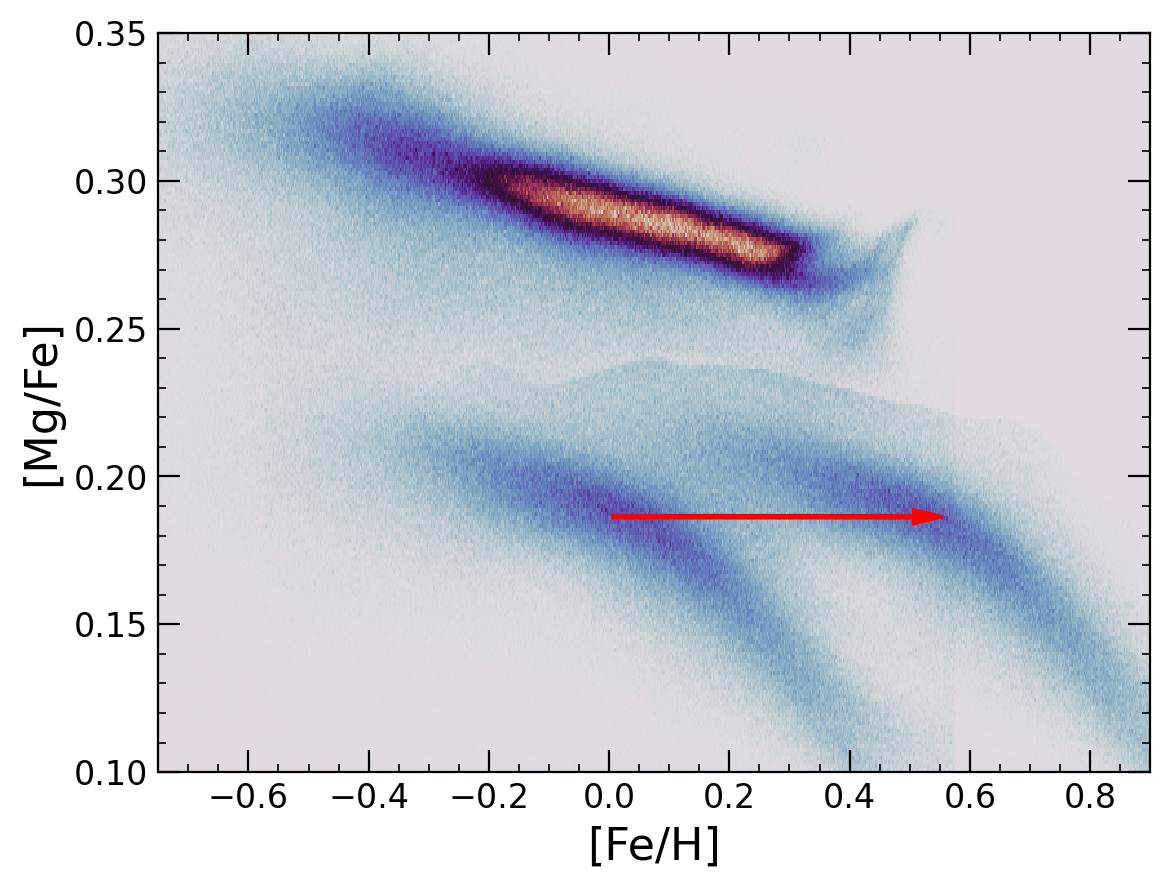}
\caption{
[Mg/Fe] versus [Fe/H] for m12q, as in Figure~\ref{fig:disks}.
Unlike our 3 other galaxies with a strong bimodality, m12q undergoes both a downward and leftward shift in this space during its $\alpha$ transition period.
To disambiguate the importance of these two shifts, we ``undo'' by hand the leftward shift to lower [Fe/H] that m12q experienced during this period, as the red arrow shows, which corresponds to the difference in gas [Fe/H] between m12q and the galaxy that merged with it (see Figure~\ref{fig:m12q}).
Both galaxies had similar [Mg/Fe] in their gas.
Shifting these low-$\alpha$ stars does not ``reconnect'' the two sequences: a bimodal gap persists.
We find that recalculating the gap width for the shifted points across the remaining overlapping region ([Fe/H]$\approx 0.2-0.5$) gives $\Delta[\rm Mg/Fe] = 0.07$, which would still be the largest value of all four galaxies (see Table~\ref{tab:galaxies}).
\textit{Therefore, as with our other 3 galaxies, the downward shift to lower [Mg/Fe] is the primary reason for the bimodality.}
That said, the leftward shift in [Fe/H] does lead to the high- and low-$\alpha$ tracks that overlap across a larger range of [Fe/H].
Therefore, we conclude that, despite the dilution in [Fe/H], m12q does \textit{not} follow the classic ``two infall'' model for the origin of a bimodality.
}
\label{fig:m12q_shift}
\end{figure}

m12q has the largest gap in [Mg/Fe] (0.11 dex) and the most complex $\alpha$ transition period.
After a $>2 \Gyr$- period of high, bursty SFR, m12q largely depleted its gas supply, declining from $f_{\rm gas} = 0.2$ $8 \Gyr$ ago to $f_{\rm gas} = 0.02$ at $6.7 \Gyr$ ago.
The resultant rise in the ratio of WDSN/CCSN rates (Figure~\ref{fig:sfr}, bottom) caused a first rise in $f_{\rm low-\alpha}$ at $6.7 \Gyr$ ago.
$6.55 \Gyr$ ago, after m12q had already begun to transition to low-$\alpha$, a massive ($M_{\rm star} \approx 2.6 \times 10^{9} \Msun$), gas-rich ($f_{\rm gas} \approx 0.84$) satellite directly merged with it (visible in Figure~\ref{fig:m12q}, first row).
This merger occurred at the same time as a temporary spike in $f_{\rm low-\alpha}$.
This brief spike arose during a brief period of highly incomplete mixing in star-forming gas, when stars formed a wide range of both [Fe/H] and [Mg/Fe].
During this brief phase, roughly half of the stars formed at high-$\alpha$ and half at low-$\alpha$.

During and immediately after this, the gas deposition and dilution from the gas-rich merger led to a significant reduction ($\approx 0.5$ dex) in both [Fe/H] and [Mg/H] (Figure~\ref{fig:alphafrac}).
Interestingly, because [Fe/H] and [Mg/H] reduced by a similar amount, the ratio [Mg/Fe] hardly changed.
As this occurred, gas began to mix more efficiently, and $f_{\rm low-\alpha}$ settled back to low, near-constant value ($0.1 - 0.2$) for the following $\approx 0.8 \Gyr$.
This is why we separate this ``first'' or ``false'' start, $t_{\rm onset,1}$, from m12q's ``actual'' $\alpha$ transition onset, $t_{\rm onset}$.

Critically important is that after the merger-induced reset in [Fe/H] (and [Mg/H]), stars continued to form primarily at high, not low, $\alpha$.
The true (permanent) $\alpha$ transition onset occurred $\approx 0.8 \Gyr$ later, as the merger-induced period of high SFR wound down, and the resultant increase in relative WDSN/CCSN rates fundamentally caused the reduction in [Mg/Fe], at relatively fixed [Fe/H], that led to a bimodality, similar to our other 3 galaxies.
Thus, while m12q experienced a key feature of the classic ``two-infall'' model \citep{Chiappini_1997}, namely a rapid reduction in [Fe/H] (and [Mg/H]) from a gas-rich merger, this occurred \textit{well before}, by $\approx 0.8 \Gyr$, the onset of the $\alpha$ transition, and it was not the primary reason for the formation of a bimodality.

To further quantify this, Figure~\ref{fig:m12q_shift} shows [Mg/Fe]-[Fe/H] for m12q, both as it is, and if we shift the low-$\alpha$ stars horizontally to the right by 0.5 dex.
Roughly, this approximates how m12q would look if it never experienced a rapid dilution in [Fe/H], for example, if the elemental abundances in the gas deposited by the merger were the same as in the preexisting gas of m12q.
Because [Mg/H] was diluted/shifted by a similar amount as [Fe/H], we apply no shift to [Mg/Fe].
The key result of Figure~\ref{fig:m12q_shift} is that even without the leftward shift in [Fe/H], m12q still would be strongly bimodal.
In this case, we measure a gap width of the $\alpha$ bimodality of $\Delta {\rm [Mg/Fe]} = 0.07$.
This is moderately weaker than $\Delta {\rm [Mg/Fe]} = 0.11$ for m12q as is, so the rapid dilution of [Fe/H] promoted a stronger bimodality, but it is a sub-dominant contribution.
Figure~\ref{fig:m12q_shift} also shows that the rapid dilution of [Fe/H] led to more overlap in the high- and low-$\alpha$ tracks of m12q, across a wider range of [Fe/H], so in this sense, more similar to the MW.

Therefore, we conclude that a rapid dilution of [Fe/H] from a gas-rich merger promoted a \textit{stronger} $\alpha$ bimodality, with more overlap in the high-$\alpha$ and low-$\alpha$ tracks, but the rapid dilution was not the primary reason for the origin of a bimodality.
Instead, the key origin of the bimodality in m12q is similar to our other 3 galaxies, which did not experience a rapid dilution of [Fe/H].

\section{Summary and discussion}
\label{sec:discussion}

\subsection{Summary of key results}

We analyzed the FIRE-2 cosmological simulations of MW-mass galaxies. Out of 16 galaxies, we identified 4 with a strong bimodality in [Mg/Fe] at fixed [Fe/H] across a large range of [Fe/H]. We examined the properties of stars belonging to the high-$\alpha$ and low-$\alpha$ populations today, and we examined the origins of these populations in terms of their star formation and gas accretion histories.
Our main results are:

\begin{itemize}
\item 4 of 16 ($\approx 25\%)$ FIRE-2 galaxies are capable of producing a strong bimodality in $\alpha$ elements, specifically, in [Mg/Fe] versus [Fe/H].
\item In all 4 galaxies, the high-$\alpha$ population corresponds to an older, radially compact thick disk, and the low-$\alpha$ population corresponds to a younger, radially extended thin disk.
\item The typical time duration of the transition from high- to low-$\alpha$ ranges from 0.3 Gyr to 1.2 Gyr, and onset times of this transition range from $5.5 - 6.5 \Gyr$ ago, likely later than the MW's transition, $\approx 10 \Gyr$ ago \citep[for example][]{SilvaAguirre_2018, Feuillet_2019}.
\item In all 4 galaxies, the bimodality transition occurred after the formation of a (thick) disk.
However, in 2 cases (m12i and m12q), the disk was destroyed and reformed during the $\alpha$ transition period.
\item The high-$\alpha$ population had similar elemental abundances at fixed age, forming in a relatively well-mixed galaxy.
However, during the transition to low-$\alpha$, the galaxy evolved relatively downward in [Mg/Fe] at fixed [Fe/H], as mono-age populations started to occupy a wide range of [Fe/H].
\item In all 4 galaxies, the transition to low $\alpha$ coincides with the formation of a radial gradient in [Fe/H].
Therefore, the onset of a strong radial gradient in metallicity may be a necessary condition to produce a bimodality, but it is not a sufficient condition, given that essentially all MW-mass FIRE-2 galaxies form negative gradients \citep{Bellardini_2022}.
\item These trends in spatial locations and elemental abundance evolution are similar for measuring stars at their current galactocentric radius and their birth radius.
This indicates that radial redistribution and dynamical heating did not play a major role in the formation of the bimodality or its correlation with the location of stars today.
\item In all 4 galaxies, the $\alpha$ transition occurred after a significant ($\approx 2 \times$) and fairly rapid (across $0.3 - 0.8 \Gyr$) decrease in SFR.
Critically, this increased the relative rates of WDSN to CCSN, which increased the rates of enrichment of Fe relative to Mg.
As a result, the enrichment of Fe nearly offset the dilution from gas accretion, leading to [Mg/Fe] declining rapidly at relatively fixed [Fe/H], which created the bimodality in all 4 cases.
\item None of these 4 galaxies had a complete cessation (gap) in their SFR history. Therefore, such a gap is \textbf{not} required for the formation of a bimodality, as previous works suggested \citep[for example][]{Lian_2020, Spitoni_2024, Beane_2025}. Instead, we find that a rapid decrease in SFR is sufficient.
\item In all 4 galaxies, the $\alpha$ transition period occurred during a period of low gas fraction ($f_{\rm gas} = 0.02 - 0.14$).
Having less gas for dilution likely enabled the rapid changes in elemental abundances necessary to achieve a bimodality gap between populations.
\item These 4 galaxies have unique merger histories; only one (m12q) involved a major merger coalescence near the $\alpha$ transition period.
The other 3 galaxies involved long-term ($>1 \Gyr$) interactions with a satellite that helped induce gas accretion.
\item One galaxy (m12q) underwent rapid deposition of gas from a gas-rich merger, which rapidly diluted [Fe/H] (and [Mg/H]) in gas.
We showed that this reset in [Fe/H] was \textbf{not} the primary factor in forming a bimodality, but it did make the bimodality stronger and occur across a wider range of [Fe/H].
\end{itemize}

\subsection{Discussion}

\subsubsection{Caveats to our results}
First, we discuss the limitations and caveats of this work. Galactic elemental abundance evolution is heavily influenced by the choice of supernova rates and yields, itself known to be poorly constrained, particularly the delay time distribution and metallicity dependence of WDSN \citep{Romano_2010, Joshi_2024, Cavichia_2024}. As discussed in \ref{subsec:FIRE}, these FIRE-2 simulations used the WDSN DTD from \cite{Mannucci_2006}, which has a strong prompt component. Compared to models with WDSN rates more spread out over time, such as \cite{Maoz_2017}, this could cause Fe production to increase more rapidly, making galaxy-wide elemental abundances more sensitive to changes in SFR. However, while this would affect the specific ``shape'' of stellar populations in [Mg/Fe]-[Fe/H] space, and the size of the gap between populations, it might not necessarily change whether the galaxy would have had a bimodality at all. This is consistent with \cite{Dubay_2024}, who found star formation history to be a more important factor that the choice of DTD, although the DTD can still affect the location of the high-$\alpha$ population. In future work, we will examine the effects of different SN rates and yield models in the FIRE simulations.


Furthermore, these FIRE-2 simulations do not include AGN feedback, and our sample size of 4 strongly bimodal galaxies out of 16 is relatively small. In the future, we plan to explore in more detail why most FIRE-2 galaxies do not form a bimodality.

\cite{Parul_2025} categorized the fiducial FIRE-2 galaxies into 3 groupings: non-bimodal, weakly bimodal, and strongly bimodal. In this work, we examined only 4 strongly bimodal cases, including 3 new strongly bimodal cases (m12r, m12i, m12q) not present in that study.
\cite{Parul_2025} found that late, Fe-poor inflows dilute the outer edges of the disk, facilitating the formation of the low-$\alpha$, low-Fe population. Expanding upon this, we find that in strongly bimodal cases, that a balance between 1) these Fe-poor inflows, and 2) Fe enrichment after a drop in SFR, leads to a decrease in [Mg/Fe] at roughly fixed [Fe/H]. This effect is also visible in moderately bimodal cases \citep{Parul_2025}.

Along with \cite{Parul_2025}, our results imply that most MW-mass galaxies would not be bimodal, which agrees with recent observations of M31 \citep{Nidever_2023}, which does not show a clear, strong bimodality like the MW. This also agrees with \cite{Mackereth_2018}, who found that around 5\% of MW-mass galaxies in the EAGLE simulation exhibit bimodalities in $\alpha$ elements. Future observations of stellar populations in nearby galaxies by \cite{vandeSande_2023} may improve our understanding of this topic.

We do not attempt to make a rigorous statement about the incidence rate in this paper, in part because \cite{Parul_2025} showed that $\alpha$ bimodalities in FIRE-2 galaxies come in degrees. In any case, we note that the incidence rate of \textit{strong} bimodalities is $\sim 25\%$, in accordance with the MW and M31, which together imply $\sim50\%$.
The consensus from cosmological simulations is that an $\alpha$ bimodality is not ubiquitous; while there is disagreement in the frequency, nearly all statistical studies find a prevalence fraction well above zero and well below 100\%, compatible with our results.

One of our key findings that holds true for all galaxies we studied is that the high-$\alpha$ population evolved ``horizontally'' (from low-Fe to high-Fe at roughly fixed [Mg/Fe]) in narrow bins of [Fe/H], and, after the $\alpha$ transition period, the low-$\alpha$ population evolved ``downwards'' (from high-[Mg/Fe] to low-[Mg/Fe] at roughly fixed [Fe/H]) across a wide range of [Fe/H]. Indeed, recent APOGEE data analysis in \cite{Cerqui_2025} shows a variety of stellar ages across the entire width of the low-$\alpha$ population, which agrees with our finding.

\subsubsection{Comparison to other works}

We now compare our results with those of other cosmological and isolated galaxy simulations.
Our results agree with the analysis of the Auriga simulations \citep{Grand_2017}, which identified a connection between the appearance of a new population in [$\alpha$/Fe]-[Fe/H] space and sudden changes in SFR from higher to lower \citep{Grand_2018, Orkney_2025}, both with and without mergers. \cite{Khoperskov_2021} also identified a similar correlation in an isolated disk simulation, and that the decrease in SFR is generally not merger-related. \cite{Beane_2025} found similar results in IllustrisTNG.
Our results agree with multiple simulations regarding the connection between the appearance of the low-$\alpha$ population and the formation of a radial gradient in [Fe/H], characterized by a low-$\alpha$ population that shows a wide range of stellar ages across its range of [Fe/H] \citep{Khoperskov_2021, Agertz_2021, Orkney_2025}.
Regarding the prevalence of this feature being relatively low among MW-mass galaxies, we agree with the analysis of the Auriga simulations \citep{Orkney_2025}, which found a prevalence rate of $\approx 50\%$, as well as the EAGLE simulations \citep{Mackereth_2018} at $\approx 5\%$. 
We note that direct comparison is difficult because we consider only ``strong'' bimodalities (as defined in Section~\ref{subsec:define}) while these sources use more inclusive definitions.

In summary, other works using cosmological simulations found overall agreement regarding the population demographics of the high- and low-$\alpha$ populations (age, geometric extent), radial migration not being necessary, and the $\alpha$ transition coinciding with a drop in SFR.
Other works found mixed agreement on the involvement of mergers, and disagreement on the timing of the transition period, with other simulations demonstrating earlier onset times.

We now compare our results with analytic models.
Our findings agree with those of \cite{Snaith_2015} regarding the SFR of bimodal galaxies: first, a period of high star formation, followed by a dip (quenching), followed by slow, sustained star formation.
\cite{Haywood_2016} shows that this scenario agrees with MW age and abundance observations.
Two of our galaxies (m12i, m12q) show a similar pattern, while the other two (m12r, m12b) are similar but lack a substantial dip.
We agree with abundance evolution tracks in \cite{Sharma_2021}, based on the model from \cite{Schonrich_2009}, that stars at all birth radii evolve together ``down'' the high-$\alpha$ region, then ``fan out'' across wide ranges of [Fe/H] in the low-$\alpha$ region, each moving downward in [Mg/Fe] at fixed [Fe/H].
However, we find these tracks without the need for radial redistribution of stars after birth to drive the $\alpha$ bimodality or significantly affect is spatial patterns today (see Figure~\ref{fig:hayden} and Figure~\ref{fig:paths}).

Our results disagree with some works, although given the possibility of the $\alpha$ bimodality having multiple possible causes, relating to different formation histories, our results do not necessarily contradict these. We do not see any evidence for a clump origin \citep[as in][]{Clarke_2019, Garver_2023}, a bar-related origin (although none of the galaxies in this study exhibit bars) \citep{Beane_2025}, or a direct merger origin \citep[as in][]{Buck_2020, Rey_2023}, although further analysis in \cite{Buck_2023} shows that the merger origin may be more indirect, as we find.

We now address the relation between our results and well-known scenarios: the two-infall model \citep{Chiappini_1997} and radial redistribution/migration \citep{Schonrich_2009}. Superficially, m12q appears similar to the two-infall model, in that a direct merger with a massive, gas-rich galaxy occurred prior to the $\alpha$ transition. However, we emphasize two counterpoints: 1) the transition did not occur immediately after; rather, $f_{\textrm{low-}\alpha}$ remained $\lesssim 0.2$ for $\sim 1$ Gyr after the merger before ``restarting'' the transition, and 2) even if we ``undo'' the addition of Fe-poor gas, as in Figure~\ref{fig:m12q_shift}, a gap in [Mg/Fe] still exists. Thus, we find that neither m12q, nor any of our other examples, follow the two-infall model. Regarding the importance of radial redistribution/migration, Figures~\ref{fig:hayden} and \ref{fig:paths} demonstrate that it has no effect on the galaxy-wide formation of a bimodality, and it is of secondary importance to the spatial distribution of this feature, relative to the patterns in stars at birth, at least in the FIRE-2 simulations.

To summarize, we broadly agree with other cosmological simulations, in that 1) most of them transition from high- to low-$\alpha$ after a sudden, rapid decrease in SFR, and 2) the transition can be related to a merger, but it is not always, and if it is, it has more to do with the long-term effects of accretion on the SFR than a difference in metallicity between host and satellite, and 3) that we expect a generally low prevalence for this feature among MW-mass galaxies.

\begin{acknowledgments}
MB and AW received support from NSF, via CAREER award AST-2045928 and grant AST-2107772.
SRL acknowledges support from NSF grant AST-2109234 \& AST-2511388 and HST grant AR-16624 from STScI.
We acknowledge HPC at UC Davis for providing computational resources that enabled our analysis.
We thank Erik Tollerud for productive and useful discussion relating to this work.
We generated FIRE-2 simulations using: XSEDE, supported by NSF grant ACI-1548562; Blue Waters, supported by the NSF; Frontera allocations AST21010 and AST20016, supported by the NSF and TACC; Pleiades, via the NASA HEC program through the NAS Division at Ames Research Center.
We acknowledge the Texas Advanced Computing Center (TACC) at The University of Texas at Austin for providing computing resources that contributed to our results.
The data in these figures, and the Python code that we used to generate them, are available at \url{https://bitbucket.org/meganbarry/bimodality_2026}.
FIRE-2 simulations are publicly available \citep{Wetzel2023, Wetzel2025} at \url{http://flathub.flatironinstitute.org/fire}.
Additional FIRE simulation data is available at \url{https://fire.northwestern.edu/data}.
A public version of the \textsc{Gizmo} code is available at \url{http://www.tapir.caltech.edu/~phopkins/Site/GIZMO.html}.
\end{acknowledgments}


\bibliography{sample701}{}
\bibliographystyle{aasjournalv7}



\end{document}